%% file: main.tex
\documentclass{article}

\usepackage{graphicx}
\usepackage{comment}
\usepackage{fullpage}
\usepackage{wrapfig}
\usepackage{amsmath, amsthm, amssymb}
\usepackage{algorithm}
\usepackage{algorithmicx,algpseudocode}
\usepackage{mathtools}
\usepackage{thm-restate}
\usepackage[numbers,sort&compress]{natbib}
\usepackage{subfig}
\usepackage{url}
\usepackage{hyperref}
\usepackage{cleveref}
\usepackage{footnote}
\makesavenoteenv{tabular}
\makesavenoteenv{table}
\usepackage{tablefootnote}

\usepackage{multirow}
\usepackage[table,xcdraw]{xcolor}
\usepackage{adjustbox}
\usepackage{fullpage}
\usepackage{pdflscape}

\usepackage{xcolor}
\definecolor{DarkGreen}{rgb}{0.1,0.5,0.1}

\graphicspath{{figuresPDF/}}

\begin{document}

\title{How we browse: Measurement and analysis of digital behavior}
\author{Yuliia Lut\footnotemark[1] \and Michael Wang\footnotemark[2] \and Elissa M. Redmiles\footnotemark[3] \and Rachel Cummings\footnotemark[1]}

\renewcommand{\thefootnote}{\fnsymbol{footnote}}
\footnotetext[1]{Department of Industrial Engineering and Operations Research, Columbia University. Emails: {\tt \{yl4737,rac2239\}@columbia.edu}. Part of this work was completed while Y.L. and R.C. were at the Georgia Institute of Technology.}
\footnotetext[2]{Email: {\tt mzywang@gmail.com}. Part of this work was completed while M.W. was at the Georgia Institute of Technology.}
\footnotetext[3]{Max Planck Institute for Software Systems. Email: {\tt eredmiles@mpi-sws.org}.
}
\renewcommand{\thefootnote}{\arabic{footnote}}

\maketitle

\begin{abstract}
Accurately analyzing and modeling online browsing behavior play a key role in understanding users and technology interactions. In this work, we design and conduct a user study to collect browsing data from 31 participants continuously for 14 days and self-reported browsing patterns. We combine self-reports and observational data to provide an up-to-date measurement study of online browsing behavior. We use these data to empirically address the following questions: (1) Do structural patterns of browsing differ across demographic groups and types of web use?, (2) Do people have correct perceptions of their behavior online?, and (3) Do people change their browsing behavior if they are aware of being observed? In response to these questions, we find significant differences in level of activity based on user age, but not based on race or gender. We also find that users have significantly different behavior on Security Concerns websites, which may enable new behavioral methods for automatic detection of security concerns online.  We find that users significantly overestimate the time they spend online, but have relatively accurate perceptions of how they spend their time online. We find no significant changes in behavior over the course of the study, which may indicate that observation had no effect on behavior, or that users were consciously aware of being observed throughout the study.
\end{abstract}

\input{intro}

\input{relatedwork}
\input{methods}

\input{rq1}

\input{rq2}
\input{rq3}

\input{discussion}

\bibliographystyle{plainnat}
\bibliography{ref1}

\newpage

\appendix

\input{appendix}

\end{document}

%% file: intro.tex
\section{Introduction}
The amount of time users spend online, and how they spend that time, has been found to relate to their digital skills~\cite{hargittai2010digital}, to the amount of social capital and other benefits they can derive from online activity~\cite{ellison2007benefits}, and to students' academic performance %
\cite{calafiore2011effect}. Thus, to draw inferences and conclusions about a variety of different digital constructs, researchers seek to measure people's digital behavior. 

While ideally researchers would be able to directly observe users' browsing behavior, due to difficulties obtaining access to such data, researchers often rely of users' self reports of their online behavior~\cite{williams2017measuring,goel2012does}. Potential concerns have been raised about the accuracy of such self-report data~\cite{redmiles2018asking,10.1145/3313831.3376435, wash2017can}. Yet, prior work measuring users' browsing behavior was all conducted using proprietary, industry data to which the majority of academic researchers do not have access (see, e.g., \cite{kumar2010characterization, goel2012does, hu2007demographic}). %

 An alternative to self-reports that are feasible in an academic research setting, are observational methods such as having participants install a browser plugin that observes and measures their behavior. Such methods are not without limitations, however. A broad literature in behavioral economics has shown that people behave differently when they are aware that their actions are observed (e.g., \cite{AB09, ABM09, HMS96, OWG92,RT04}). However, this literature has focused primarily on behavior in incentivized economic games, not web behavior.  %
 
In this work, we apply and compare both self-report and observational methods to provide an up-to-date --- the most recent measurement study of online browsing behavior, to our knowledge, was conducted in 2013~\cite{abramson2013associative} --- understanding of both user's browsing behavior and a comparison of two academically-feasible methods for measuring this behavior. To do so, we designed and conducted a user experiment ($n=31$) in which we both surveyed participants about their browsing behavior and observed participants' browsing behavior continuously for 14 days. %
Using these data, we address the following research questions: %

\begin{itemize}
\item[(RQ1)] Does browsing behavior differ across user groups (i.e., demographics) and types of web use? 
\item[(RQ2)] Do people have accurate perceptions of their behavior online? Does perception accuracy differ by user group or type of web use?
\item[(RQ3)] Do people change their browsing behavior if they are aware of being observed? 
\end{itemize}

For RQ1, we observe that people spend much more time online, relative to prior work conducted in 2010:  median of 2.9 hours per day in our study versus one hour per day in \cite{kumar2010characterization}. We found little difference across demographic groups by race and gender, but did find significant differences by age, with older participants (aged 35-44) browsing less than younger groups (aged 18-24 and 25-34) across multiple metrics of browsing activity. We find few significant differences in within-website browsing behavior across different categories of websites. One notable exception is the Security Concerns category, which had significantly different within-website browsing patterns than all other categories ($p<10^{-5})$. We suggest ways that this finding can be used to automate detection of security concerns online. %

For RQ2, we find that people substantially overestimate their time spent online (80.6\% of our participants, by an average overestimate of 4.5 hours). This overestimation effect persists, even after controlling for various methodological alternatives. We find no significant difference across demographic groups, meaning that all groups overestimate their time spent online. We also find that people have roughly accurate perceptions of their top-browsed website categories: 50.3\% of reported top browsing categories were indeed in the participant's observed top browsing categories.

We are unable to directly test RQ3 because we obtained informed consent for data collection from all participants in our study. Instead, we test whether behavior changed over time during the study, under the hypothesis that participants will have higher awareness of observation early in the study, shortly after providing consent, and lower awareness later in the study. We do not find changes in either level of browsing activity or in distribution of browsing across website categories, over time during the study. This could indicate that people do not change their behavior when aware of being observed, or it could be that a 14-day study is not sufficient time for participants to forget that their browsing data are being collected.

%% file: relatedwork.tex
\section{Related Work}
\label{s.relatedwork}
In this section, we review prior work on users' browsing behavior and perceptions of that browsing behavior.

\paragraph{Browsing Behavior Measurements.} Prior work finds correlations between users' browsing behavior and their demographic or behavioral type.  \citet{goel2012does} use large-scale measurement data to study the differences in how various demographic groups use the internet, and show that use of different website types depends more on users' level of education than on their demographic features. 
\citet{kumar2010characterization} propose taxonomy of page views for popular website categories and study the behavior of Yahoo! users based on search and toolbar log data. Their analysis includes website categorization but does not include demographic data. They show that the distribution of page views across website categories is skewed, with the top five categories  (news,  portals,  games, verticals,  multimedia) accounting for more than half of all Web activity.  %
\citet{hu2007demographic} offer a methodology to predict certain demographic features such as gender and age from a user's observed browsing behavior, which provides 30.4\%  and  50.3\% improvements  on  gender and age  prediction  respectively compared to baseline algorithms. %
\citet{abramson2013associative} show that a user's pattern of web browsing behavior can be uniquely identified by the types of websites they access and the time-of-day they access those websites with at least 75\% accuracy.  %

While these works lay a foundation for measuring user behavior online, the most recent work in this literature \cite{abramson2013associative} is nearly a decade old. A more updated understanding of browsing behavior under modern internet usage is needed. Additionally, these works measure user browsing behavior, but do not elicit user perceptions of their own browsing behavior. In this work, we both elicit self-perceptions of browsing and measure browsing behavior, which allows us to evaluate accuracy of users' perceptions.

\paragraph{Accuracy of user perceptions.} The accuracy of user perceptions of browsing behavior has been previously studied in limited contexts. \citet{10.1145/3313831.3376435} study how accurately users estimate their time spent on Facebook. They show that self-reported data can be often unreliable, and that people tend to overestimate the time they spend on Facebook but underestimate the number of times they visit. \citet{calafiore2011effect} study dependencies between academic grades and the time students spend on an online educational platform, and find that longer times spent online are associated with higher grades.
We extend this work in two key ways. First, we examine the accuracy of users' perceptions of their overall browsing behavior rather than their behavior on one specific website. Second, we test not only the accuracy of users' perceptions about the time they spend online, but also which categories of websites they most frequently browse.

Outside of studies of online browsing behavior, prior work in the security domain has examined the accuracy of people's self reports. \citet{wash2017can} follow a methodology similar to our own, comparing observed behavior via software that participants installed and which logged their behavior over six weeks to self-reports from those same participants about their digital security behavior. They find low correlation between participants' self-reported and actual behavior across a majority of behaviors that were observed. \citet{redmiles2018asking} specifically examine software updating behavior, comparing self-reported behavioral intentions in response to software update prompts with behavioral responses to the same prompts observed through proprietary industry data. They find a significant correlation between responses, but find that self-reporting participants reported that they would update significantly faster than observed users did.

%% file: methods.tex
\section{Methods}
To answer our research questions we observed the browsing behavior of 31 participants over a period of 14 days in August and September 2019, and additionally assessed participants' self-reported perceptions of their browsing behavior. In this section we describe our study procedures (Section \ref{s.procedures}), data collection from both the Chrome extension and self-reported data (Section \ref{s.data}), data analysis (Section \ref{s.analysis}), and the limitations of our work (Section \ref{s.lim}). All study procedures were approved by the Georgia Institute of Technology's Institutional Review Board (IRB).

\subsection{Study Procedures}\label{s.procedures}

We recruited participants by advertising flyers on bulletin boards and student gathering spaces on the campus of the Georgia Institute of Technology, which is a large public institution for higher education.  The flyers advertised an ``Internet Browsing Study'' stating: ``The purpose of this study is to determine how real people interact with the internet so that we can better protect user data.'' The flyer also included a link to the online screening survey where participants could verify eligibility and sign up for the study. This flyer is shown in Figure \ref{fig:flyer} in Appendix \ref{app.screenshots}. 

The brief screening survey was hosted on Qualtrics, and verified that participants met the eligibility criteria for our study: participants needed to be aged 18 or older, native English speakers, and needed to browse the internet at least 5 hours per week. The age requirement ensured we did not have any minors in our study; the English language requirement ensured that our collected data would focus primarily on English-language websites; and the browsing activity requirement ensured that our study participants would generate sufficient browsing data during the study.

Participants who met the eligibility criteria were invited to come into the lab to complete a consent form for the experimental portion of the study, complete a pre-study survey, and install the Chrome extension that would collect their browsing data. We asked participants to come in-person so that we could provide support in installing the extension, as an effort to mitigate issues of digital inequity. The pre-study survey asked participants to self-report their demographic information and perceptions of their own browsing habits. 

Over the next 14 days, the extension collected data on participants' web browsing, including their websites visits, actions within each website, and timestamps of each browsing action. Finally, on the 14th day of the study, participants returned to the lab, where they uninstalled the Chrome extension and completed a brief post-study survey that asked whether participants changed their browsing behavior over the course of the study. For those individuals who were not able to return to the lab in-person on their 14th day, we truncated data collection after 14 days. More details on the pre-study survey, post-study survey, and extension-based data collection are all given in Section \ref{s.data} below.

Participants were paid \$200 for their full participation in the study. Participants had the option of exiting the study early and receiving payment proportional to the length of their participation. No participants exercised this option.

\subsection{Data Collection}\label{s.data}

In this section we describe the data that were collected in our study. Data from the browsing extension are described first in Section \ref{s.extdata}, and then survey data are summarized in Section \ref{s.surveydata}.


\subsubsection{Extension-based Data} \label{s.extdata}
In order to record participants' browsing behavior and the metadata related to it, we developed a system that includes a Chrome browser extension for data collection and a server where the collected data are sent and stored.
The extension monitors events in a browser using scripts in its background service worker. 
We chose a set of user browsing actions to observe through this extension, which included: hitting the back button or forward button ({\it backButton}), creating a new tab either manually or by opening a link in a new tab
({\it newTab}), changing tabs ({\it tabChange}), typing in the address bar ({\it omniBox}), going to a new URL either by using the address bar or clicking a link in a page ({\it urlChange}), clicking a button in a webpage that does not change the URL (e.g., `Like' on social media) ({\it click}), and typing in a textbox within a webpage ({\it type}). We chose these because they are common browsing actions that generate or affect internet packets from a user's browsing activity. Note that some of these actions occur within a fixed webpage ({\it urlChange, click, type}), while others are not necessarily affiliated with a specific website ({\it backButton, newTab, omniBox, tabChange}).
We added an additional {\it awake} action which the extension would generate every 5 minutes if the browser was open and their computer was connected to the internet. This signal was intended to verify that participants had not uninstalled or disabled the extension during the study. No participants were removed from the analysis from missing {\it awake} actions. These observable events are summarized in Table \ref{tab:action-types-class}. 

\begin{table}[tbh]
\small
\centering
\begin{tabular}{|l|l|}
\hline
awake      & \begin{tabular}[c]{@{}l@{}}
This action indicates that a user is online. Appears \\
every 5 minutes when browser is open and online.\\ Can occur when a user is not actively
browsing. 
\end{tabular} 

\\ \hline
backButton & 
Clicking on the back button                                                                                                                                                                                                                                         \\ \hline
click      & \begin{tabular}[c]{@{}l@{}}
Click that does not cause URL change.
\end{tabular}                                                                                                                                                                                                 \\ \hline
newTab     & \begin{tabular}[c]{@{}l@{}}
Opening new tab                                                       \end{tabular}                                                                                                                                                \\ \hline
omnibox    & \begin{tabular}[c]{@{}l@{}}
Typing in omnibox (address bar / search engine)    
\end{tabular}                                                                                                                                                                                   \\ \hline
tabChange  & \begin{tabular}[c]{@{}l@{}}
Alternating between existing tabs     \end{tabular}                                                                                                                                                                                                                                 \\ \hline
type       & Typing a single character                                                                                                                                                                                                                                                             \\ \hline
urlChange  & \begin{tabular}[c]{@{}l@{}}
Click that causes URL change. 
\end{tabular}                                                                                                                          \\ \hline
\end{tabular}

\caption{Action types collected through the extension}
\label{tab:action-types-class}
\end{table}

When one of these actions occurred in a user's browsing, the Chrome extension recorded the action and relevant metadata including the time the event occurred, the URL (if any) on which the action occurred, and participant performing this action, and forwarded these data to a secure server.

To ensure privacy of the participants, each one was assigned a random ID that was used to associate them with their browsing actions.  In this way, we could track the actions of each participant without linking these data to his or her identity.  The pre- and post-study surveys described next were also associated with participants' random IDs, rather than their names or other identifiers. However, it is still possible that the URLs of visited websites may be disclosive, particularly for long URLs that embed information beyond the domain name. To protect participants, we truncated the URLs in our collected data to contain only the domain name of the website that was visited, and removed the subdirectory information. For example, {\tt www.facebook.com/UserName} became {\tt www.facebook.com}. This did not affect our analysis because we are still able to track URL changes within a fixed domain name with the {\it urlChange} action.

To enable analysis of patterns of web use, we categorized the websites browsed by participants using the Symantec WebPulse Site Review tool \cite{Symantec}.  This tool offers three levels of website categorization: categories, subgroups, groups of categories. For this work, we focus on subgroups of categories because they give the right level of granularity for our analysis: groups are too broad and not informative, while categories are too narrow and do not allow for statistical significance of tests due to the large number of categories. For ease of presentation, we refer to the subgroups simply as ``categories'', since these are our unit of measure for website categorization. These categories -- along with examples of each -- are presented in Table  \ref{tab:categorization}.

\begin{table}[tbh]
\centering
\small
\begin{tabular}{|l|l|}
\hline
\textbf{Category}   & \textbf{Subcategory (examples)} \\ \hline
Adult Related  &  Adult/Mature Content,   Gore/Extreme  \\ \hline
Liability Concerns &  Piracy/Copyright Concerns, Violence/Intolerance \\ \hline
Security Threats   &  Malicious Outbound Data/Botnets, Phishing  \\ \hline
Security Concerns   & Compromised Sites,  Hacking, Spam   \\ \hline
File Transfer   & File Storage/Sharing, Peer-to-Peer (P2P) \\ \hline
Society/Government  & Charitable/Non-Profit,  Government/Legal \\ \hline
Social Interaction  & Personal Sites, Social Networking \\ \hline
Multimedia  &  Audio/Video Clips, Media Sharing \\ \hline
Communication & Email, Internet Telephony,  Online Meetings \\ \hline
Health Related  &  Health, Restaurants/Food, Tobacco \\ \hline
Leisure   & Art/Culture, Entertainment, Games \\ \hline
Commerce   &  Cryptocurrency, Job Search/Careers, Shopping \\ \hline
Technology  & Cloud Infrastructure, Computer/Information Security \\ \hline
Information Related & Education, News, Reference, Search Engines/Portals \\ \hline
\end{tabular}
\caption{Website categories, Symantec WebPulse Site Review \cite{Symantec}}
\label{tab:categorization}
\end{table}

The  JavaScript code for the extension, along with a description, can be found at \linebreak \url{https://github.com/mzywang/browsing-experiment-extension}. This code can be used for replication of our study, and may be of independent interest for future research involving browsing data collection.

\subsubsection{Self-reported Data}\label{s.surveydata}
Two surveys were used to assess participants' self-perceptions of their browsing behavior and to collect demographic data. They were conducted immediately before and after the period of browsing data collection.

In the pre-study survey we asked participants to report their age, gender, ethnicity, and race. We also asked participants to report ``How many hours per day, on average, would you say that you spend online?'', reported on a slider from 0 to 24 hours; and ``What are your most frequented categories of websites to visit? Please select all that apply.'', with answer choices: ``Social Network (Facebook, Instagram, Reddit, etc)'', ``Business (Onenote, Dropbox, Linkedin, etc)'', ``Entertainment (Youtube, Netflix, IMDB, etc)'', ``News (CNN, ESPN, etc)'', ``Search (Google, Bing, etc)'', ``Banking (Paypal, Any personal bank, etc)'', ``Shopping (Amazon, Walmart, etc)'', ``Blogging (Tumblr, Wordpress, etc)'', ``Reference (Wikipedia, Weather, etc)''. Additional questions related to internet use and identity were asked, but are not analyzed in this work.%

In the post-study survey, we asked participants whether: ``During the course of the study, did you change your browsing behavior to prevent information from being learned about you?'', with answer choices ``Yes'' and ``No''. (As in the pre-study survey, additional questions related to internet use were asked but are not analyzed in this work.)

\begin{table}[tbh]
\small
\centering
\begin{tabular}{|l|l|l|}
\hline
 {\bf Demographic}                               & {\bf Group  }         &  {\bf Number (\%)}  \\ \hline
\multirow{2}{*}{Gender}            & Female          & 12 (38.7\%) \\
                                & Male            &  19 (61.3\%)	\\\hline
\multirow{3}{*}{Age}          & 18-24           &14  (45.2\%) \\
                                & 25-34           & 14 (45.2\%) \\
                                & 35-44           & 3 (9.7\%)  \\ \hline
\multirow{4}{*}{Race\footnotemark }          & Asian           & 7 (22.6\%)  \\
                                & Black or African American          & 10 (32.3\%) \\
                                & White           & 9 (29.0\%)  \\ 
                                & Two or more races    &3 (9.7\%)\\
                                \hline                               
\multirow{2}{*}{Nationality}        
								& USA             & 16 (51.6\%) \\
                                & Other           & 15 (48.4\%) 	
     \\\hline
\end{tabular}
\caption{Participant demographics }
\label{tab:sample-demo-tab}
\end{table}

Our participants were primarily undergraduate and graduate students at the Georgia Institute of Technology. Demographics features of our participants are presented in Table \ref{tab:sample-demo-tab}. 
\footnotetext[1]{One participant preferred not to disclose their race, and one participant responded with their ethnicity instead of race.}

\subsection{Data Analysis}\label{s.analysis}

\textbf{RQ1: Differences in behavior.} We first examine differences in web use between participants of different genders, races, and ages, using two metrics of browsing activity: time spent browsing and number of browsing actions. To compute the amount of time participants spent browsing, we convert sequences of instantaneous browsing actions into \emph{clickstreams} of consecutive actions performed by one participant, which represent a period of continuous active browsing. Prior work ended a clickstream after periods of inactivity ranging from 30 seconds on Facebook \cite{10.1145/3313831.3376435} to 30 minutes across all websites.  We chose to use 30 minutes of inactivity as a cutoff because we considered the full range of internet browsing.

We use a one-sided $t$-test with the null hypothesis of no difference between the mean number of daily browsing actions (or mean number of hours spent browsing) on average across days with observed online behavior, for each relevant pair of demographic groups. For brevity, we refer to these two metrics respectively as ``daily average number of browsing actions'' and ``daily average browsing time,'' as we use these same metrics when measuring browsing activity for the other RQs.
We apply bootstrapping to account for smaller sample sizes, and we correct for multiple testing using Bonferonni-Holm correction. For statistical significance reasons, for race we compare only Asian, Black or African American, and white groups, since these have sufficient representation in our sample.

To investigate patterns of web use, we test whether participants had similar distribution of browsing actions within websites of different categories. Only three types of browsing actions could occur within a website: \emph{click}, \emph{type}, and \emph{urlChange}.\footnote{All other actions involved changing websites or actions outside of a website, such as creating a new tab, and thus could not be associated with a particular website category.} We measure the empirical distribution of these actions across category, and use a Pearson's $\chi^2$-test for homogeneity to test whether the distribution of actions were the same (pairwise) across categories.

\textbf{RQ2: Accuracy of perceptions.} To address RQ2, we compare participants' observed browsing behavior with their self-reported daily time spent browsing and most browsed website categories.  To measure differences in terms of time spent browsing, we introduce a value $\delta_i$ for each participant $i$, which is defined as the difference between their daily average browsing time, and their self-reported daily time spent online. That is, if participant $i$ spent $S_i$ total hours of active browsing across $n_i$ days of the study (i.e., they were active for $n_i$ out of the 14 days), and they self-report spending $t_i$ hours per day online, then $\delta_i$ is defined as $\delta_i = \frac{S_i}{n_i} - t_i$. We use a $t$-test to determine whether the mean of the $\delta_i$s among participants in our study is significantly different from 0, to determine whether there is a significant difference in observed and perceived browsing time.  %
We additionally examine whether differences between observed and perceived browsing time vary based on demographic group. We use a $t$-test for mean equality of the $\delta_i$s between demographic groups, and apply bootstrap techniques to improve the statistical power of our small sample.

We then examine whether participants' perceptions of their most commonly browsed types of websites are correct. Our pre-study survey asked participants to select any number of website categories that they most frequently browse. For participant $i$ who reported $k_i$ top browsing categories, we compare their reports to their top $k_i$ categories of observed browsing time. We report true/false positive/negative rates for each website category, as well as overall percentage of participants with correct/incorrect perceptions. To determine browsing time in each category, we separate clickstream browsing time by category, by dividing browsing sessions when the participant began browsing a URL of another category. For this portion of the analysis, we use categorization from Alexa Top Websites \cite{alexa} because our pre-study survey listed these category choices. Unfortunately, this Alexa Top Websites tool was retired after our study was conducted. Thus we use Alexa Top Websites categorization here for consistency with the survey, and we use the Symantec WebPulse Site Review Request for the remainder of analysis in the paper to enable reproducibility.

\textbf{RQ3: Changes with observation.} With RQ3, we aim to test whether people behave differently online when they are consciously aware of being observed.  We hypothesize that participants’ conscious awareness of being observed may be the most salient early in the study, shortly after they provide informed consent for data collection, and that this awareness may diminish over time.  This is consistent with evidence that the behavioral effects of observation are amplified by interpersonal reminders of the observation \cite{RT04, OWG92}. If this were the case, we would expect to observe a change in participants’ behavior over time during the study. As a proxy measure for this analysis, we test whether the distribution of participants’ activity during the first half of the study (Days 1-7) is different from the second half (Days 8-14).

We first test for changes in participants' level of browsing activity, as measured by both the daily average number of browsing actions and daily average browsing time. For both activity metrics, we test for changes in the mean and the variance in participants' level of browsing activity, using a $t$-test and Levene test, respectively.  We then test for changes in the distribution of website categories browsed under both metrics. We use a Wilcoxon signed-rank test to evaluate if the aggregated distribution of activity across categories from the first and the second halves of the study are different.

\subsection{Limitations}\label{s.lim}

As with any user study, our findings are subject to multiple practical limitations. First, our sample population was relatively small and consisted primarily of undergraduate and graduate students. We used bootstrapping, which is a robust and commonly-used technique for calculating estimators when sample sizes are small or assumptions about normality of the sampling distribution cannot be made~\cite{efron1986bootstrap}. Our sample is not fully representative of the internet-using population, and our results should be interpreted in this context.

Second, our surveys relied on self-reported user data. While one of the goals of this work was to measure whether users' perceptions of their browsing behavior were accurate, certain self-reported data (e.g., demographics) were unverifiable in the study. Participants could also have misinterpreted the survey questions, or changed their answers due to desirability bias towards a more acceptable behavior~\cite{groves2011survey}.

Third, users could turn off the data collecting extension at any point during the study. While this feature was necessary for ethical data collection --- participants must have the option to opt out of the study at any time --- it also allowed for a potential bias in the collected data, as participants could disable data collection on embarrassing or sensitive websites. %

Finally, our metric of active browsing time converts a series of instantaneous events into an aggregate measure of time spent browsing, as is the convention in  prior work \cite{10.1145/3313831.3376435,calafiore2011effect, abramson2013associative}. This approach does not capture passive browsing activities, such as streaming a movie or reading an article, although we do explore alternative methodologies for measuring active browsing time in Appendix \ref{app.analysis}.

%% file: rq1.tex
\section{RQ1: Does browsing behavior differ across user demographic groups and type of web use?}

In this section, we measure our participants' browsing behavior in terms of time spent browsing and number of browsing actions. We test whether that behavior differs based on the type of user (i.e.,  demographics) in Section \ref{s.demo} and type of web use (i.e., website category) in Section \ref{s.categories}. %

\begin{figure}[tbh]
	\centering
	\includegraphics[width=.65\columnwidth]{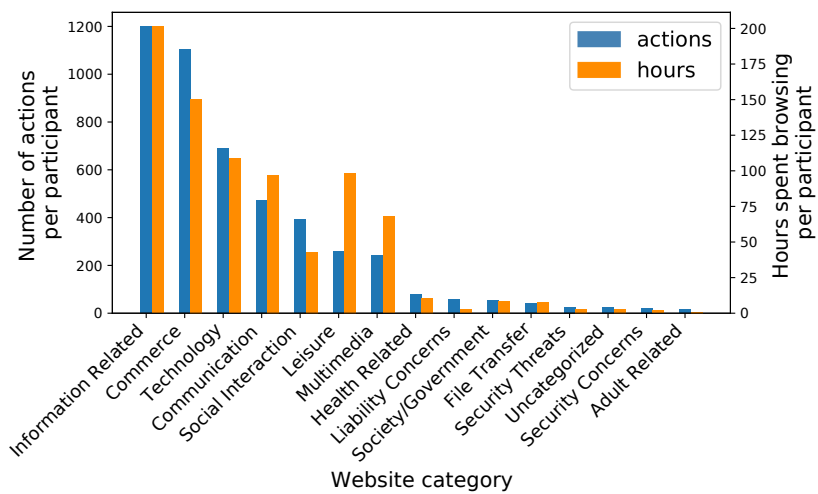}
\caption{Distribution of number of browsing actions (blue) and browsing time in hours (orange) on different website categories averaged over all participants in our study. 
} \label{fig:time_psub}
\end{figure}

Overall, we observe that our participants spent an average of 146 minutes (SD = 100.5) browsing daily during the course of the study.  Participants averaged 968 browsing actions per day (SD = 1529). Figure \ref{fig:time_psub} illustrates the distribution of time participants spent on each website category and number of browsing actions in each category.  We see that ``Information Related'', ``Commerce'', and ``Technology'' websites are the most popular according to both metrics of activity. Some categories, such as ``Social Interaction,'' ``Leisure,'' and ``Multimedia'', are popular under one metric, but not the other. This suggests, for example, that browsing ``Leisure'' and ``Multimedia'' websites does not involve as many click actions as other categories of websites. Recall the descriptions and examples of each website category given in Table \ref{tab:categorization}. Figure \ref{fig:old_hist_url} in Appendix \ref{app.figntab} shows the distribution of browsing actions and website category for the top 100 most browsed websites by all users during the study.

Compared to prior work conducted 6 and 9 years prior to our study, respectively, we observe that our participants visit similar numbers of pages per session  (5-151 per session vs. 14-130 in \cite{abramson2013associative}), but spend more time online (median of 2.9 hours per day vs. a median of an hour per day in \cite{kumar2010characterization}). 

\subsection{Differences across demographic groups}\label{s.demo}

First, we consider the differences in participants' behavior across demographic groups, motivated by prior work showing relationships between browsing behavior and demographic features \cite{goel2012does,hu2007demographic}. We explore this by testing for differences in
daily average number of browsing actions and daily average browsing time (per person) across demographic groups. 

With daily average number of browsing actions, we find no significant difference between genders ($t=-0.228$, $p=0.822$), between Black or African American and Asian participants ($t=0.688$, $p=0.502$),  between white and Asian participants ($t=-0.321$, $p=0.754$), and between white and Black or African American participants ($t=-0.760$, $p=0.461$). %
We also observe no significant difference in daily activity level between those aged 18-24 and those aged 25-34 ($t=0.322$, $p=0.999$). However, the daily activity of older participants (aged 35-44) is significantly lower than that of those aged 18-24 years and 25-34 years ($t=3.301$, $p=0.007$ and $t=2.994$, $p=0.051$, respectively). 

With daily average browsing time, we find that there is no significant difference between genders ($t=-0.073$, $p=0.950$) and among races (pairwise, $t=-0.842$, $p=0.381$; $t=-0.642$, $p=0.501$; $t=0.114$, $p=0.918$). We do not observe a significant difference in daily average browsing time between those 18-24 years old and those aged 25-34 ($t=-0.467$, $p=0.999$). However, participants aged 35-44 on average spend significantly less time online daily than younger participants ($t=3.297$, $p=0.007$ in comparison to those aged 18-24, and $t=3.187$,  $p=0.013$ in comparison to those aged 25-34, respectively). See Table \ref{tab:table_pval_demo_pair_ttest_actions} and  \ref{tab:table_pval_demo_pair_ttest_time} below for a full presentation of these tests and their $p$-values. %

\begin{table}[tbh]
    \begin{minipage}{.45\linewidth}

\small{
\begin{tabular}{|llc|}
\hline
\multicolumn{2}{|l}{Feature}                                                      & $p$-value                              \\ \hline
\multicolumn{2}{|l}{\textbf{Gender}}                                              &                                      \\
 & Male vs Female                                                                & 0.822                                  \\ \hline
\multicolumn{2}{|l}{\textbf{Race}}                                                &                                      \\
 & Asian vs Black or African American & 0.502                                  \\
 & Asian vs White                                                                & 0.754                                  \\
 & Black or African American vs White  & 0.461                                  \\ \hline
\multicolumn{2}{|l}{\textbf{Age}}                                                 &                                      \\ 
 & 18-24 vs 25-34                                                                & 0.999                                  \\
 & 18-24 vs 35-44                                                                & \bf{0.007} \\
 & 25-34 vs 35-44                                                                & \bf{0.051} \\ \hline
\end{tabular}}
      \caption{Test for equality of means of daily average number of browsing actions}
      \label{tab:table_pval_demo_pair_ttest_actions}
    \end{minipage}%
    \hfill
    \begin{minipage}{.45\linewidth}
\small{
\begin{tabular}{|llc|}
\hline
\multicolumn{2}{|l}{Feature}                                                      & $p$-value                              \\ \hline
\multicolumn{2}{|l}{\textbf{Gender}}                                              &                                      \\
 & Male vs Female                                                                & 0.950                                  \\ \hline
\multicolumn{2}{|l}{\textbf{Race}}                                                &                                      \\
 & Asian vs Black or African American & 0.381                                  \\
 & Asian vs White                                                                & 0.501                                  \\
 & Black or African American vs White  & 0.918                                 \\ \hline
\multicolumn{2}{|l}{\textbf{Age}}                                                 &                                      \\ 
 & 18-24 vs 25-34                                                                & 0.999                                  \\
 & 18-24 vs 35-44                                                                & \bf{0.007} \\
 & 25-34 vs 35-44                                                                & \bf{0.013} \\ \hline
\end{tabular}
}
 \caption{Test for equality of means of daily average browsing time.}
 \label{tab:table_pval_demo_pair_ttest_time}
    \end{minipage} 

\end{table}
 
\subsection{Differences in behavior across types of web use}\label{s.categories}
Next we explore how users' behavior within a website changes across different categories of websites. This is motivated in part by existing literature showing that users interact differently with different websites \cite{kumar2010characterization, abramson2013associative}. We aim to understand whether this behavior varies structurally by website category. We measure behavior by the distribution of browsing actions within each website category, rather than total number of browsing actions or time spent browsing because our goal is to measure differences in how users interact with a website, rather than their level of interaction.

\begin{figure}[tbh]
\centering
	\includegraphics[width=.6\columnwidth]{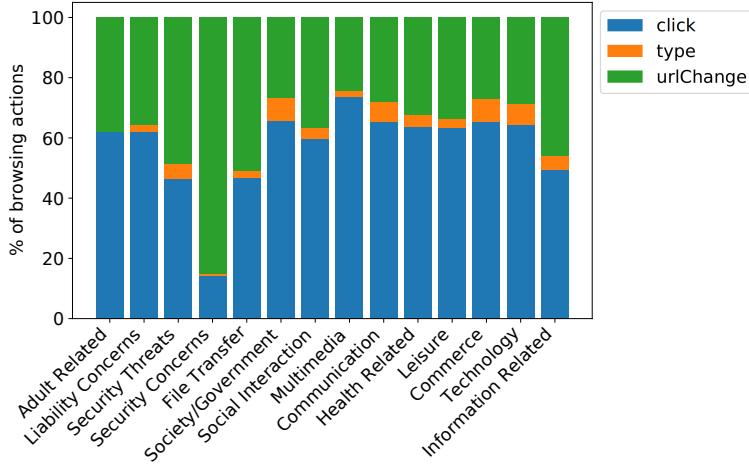}
	\\
\caption{Empirical distributions of participants' \textit{click}, \textit{type}, and \textit{urlChange} browsing actions within each website category. 
} \label{fig: actions_inside_site}
\end{figure}

Figure \ref{fig: actions_inside_site} shows the empirical distribution of click events on different website categories. We observe qualitatively that for most website categories, participant actions were mostly \emph{click}, slightly fewer \emph{urlChange}s, and a small number of \emph{type} actions.  A notable exception is Security Concerns websites, which saw significantly different behavior from all categories ($\chi^2>27$, $p<10^{-5}$ for all categories). Behavior on Multimedia websites was found to be significantly different from behavior on File Transfer ($\chi^2=15.349$, $p=0.036$) and Security Threats ($\chi^2=15.969$, $p=0.027$) websites. The observed differences between all other pairs of websites were not significant. Table \ref{tab:p-values} in Appendix \ref{app.figntab} shows the $p$-values from this test, presented for each pair of website categories.

\textbf{Security Concerns.}
In our study, 10 out of 31 participants visited a total of 62 Security Concerns websites during the two-week period during which they were observed. There were four Security Concerns subcategories that were visited by these participants: Suspicious (e.g., \texttt{www.netlflix.com}), Placeholders (e.g., \texttt{www.canvas.com}, \texttt{www.richvideos.com}), Potentially Unwanted Software (e.g., \texttt{www2.securybrowse.com}), and Hacking (e.g., \texttt{www.recoverlostpassword.com}). The majority of the Security Concerns websites visited in our study were in the Suspicious subcategory (48 out of 62).

The fact that  user behavior on Security Concerns websites differs from all other types of browsing makes sense given known typical behavior of malicious websites, which aim to redirect users to further malicious pages and/or capture their credentials~\cite{thomas2011design}. The behavior-distribution signals we observe may serve to augment existing approaches to detecting new or unknown Security Concerns websites~\cite{cao2014behavioral,cao2015detecting,dong2008user}. Additionally, such signals may be useful for developing just-in-time in-browser warnings about a potential security concern based on observed browsing behavior on the website.

%% file: rq2.tex
\section{RQ2: Do people have correct perceptions of their behavior online?}

In this section we test whether participants had accurate perceptions of their online browsing behavior. We first evaluate participants' perceptions of their time spent online and how this varies by demographic group in Section \ref{s.pertime}, and then we measure participants' perceptions of the website categories that they most frequently browse in Section \ref{s.percat}.

Overall, we observe that our participants think they spend on average 6.87 hours (SD = 4.6) per day browsing. In response to the question,  ``What are your most frequented categories of websites to visit?'', the most common answers were ``Entartainment'', ``Search'', and ``Social Network'' (respectively from 27, 27, and 23 participants out of 31). The full list of categories with the number of participants who chose each category as their most frequently visited can be found in Table \ref{tab:oldcat_users_count} in Appendix \ref{app.figntab}. 

\subsection{Perceptions of time spent browsing}\label{s.pertime}
\begin{figure}[H]
	\centering
\subfloat[][\small Observed vs perceived browsing time]{\includegraphics[width=.45\columnwidth]{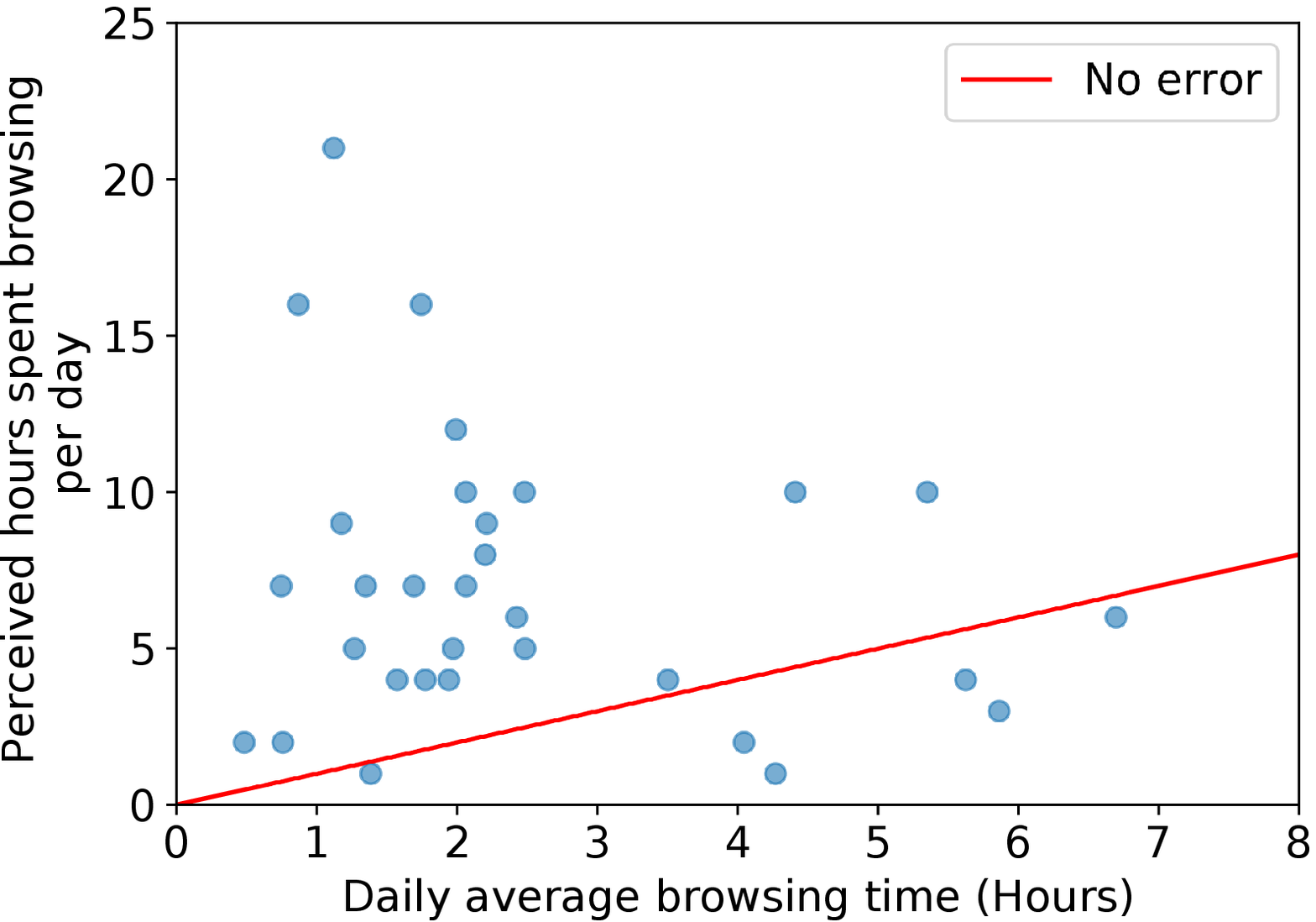}}
\subfloat[][\small Distribution of error in perceptions]{\includegraphics[width=.45\columnwidth]{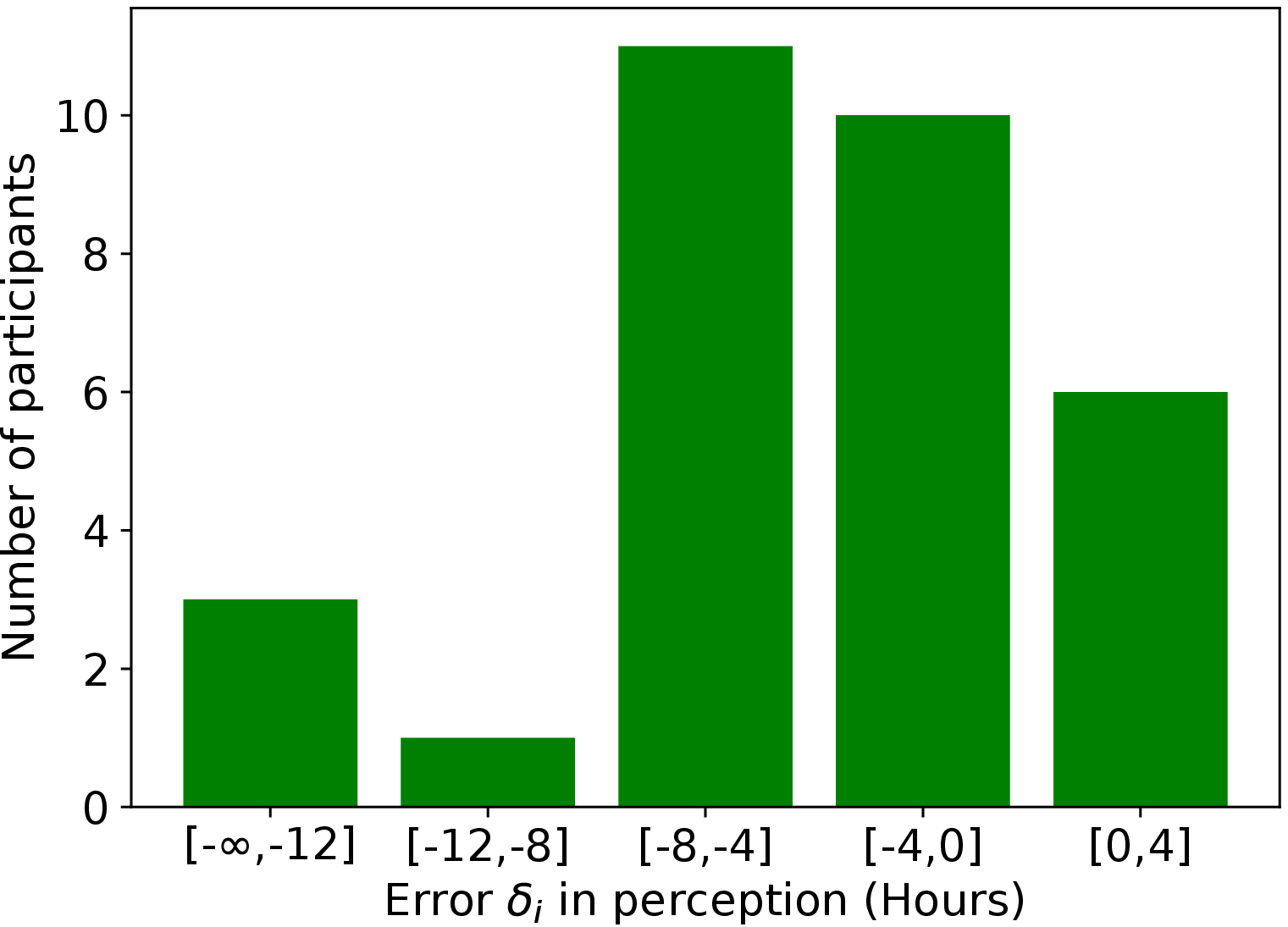}}

\caption{
(a) Scatter plot illustrating actual daily average browsing time vs. perceived (self-reported) number of hours spent browsing per day. Each point corresponds to one participant.
(b) Distribution of error values $\delta_i$ in the participant population. 
} \label{fig:delta}
\end{figure}

We find that the majority of participants (26 out of 31, 80.6\%) significantly over-reported their daily browsing time. Figure \ref{fig:delta} illustrates the relationship between participants' observed time spent browsing and their perceived (self-reported) time spent browsing. Figure \ref{fig:delta}a shows a scatter plot with one dot corresponding to each participant, where the $x$-coordinate is their daily average browsing time, and the $y$-coordinate is the self-reported daily time spent browsing. The red line ($x=y$) corresponds to no error in perceptions, and points further from this line have larger error between perceptions and actual browsing behavior. We observe that most participants substantially overestimated their time spent browsing, as evidenced by the number of points above the red line.

Figure \ref{fig:delta}b aggregates this information to illustrate the error in participants' perceptions of their browsing behavior at the population-level. Recall that $\delta_i$ is the difference between the actual (observed) daily average browsing time of participant $i$, and their self-reported (perceived) daily browsing time. Since a large fraction of participants had a negative value of $\delta_i$, we see that most participants over-reported their time spent browsing. The average error $\delta_i$ among our participants is -4.5 hours  (SD=5.24). A more detailed visualization of this error at the participant-level is illustrated in Figure \ref{fig:delta_hours_pday_puser} in Appendix \ref{app.figntab}.

\paragraph{Alternative measures of activity.}
For measurements of active browsing time, we followed the convention of \citet{abramson2013associative} to assume that a browsing session ends after 30 minutes of inactivity (i.e., no browsing actions aside from the \emph{awake} action were recorded by the browser extension for 30 minutes). Other existing literature used cutoffs ranging from 30 seconds \cite{10.1145/3313831.3376435} to 20 minutes \cite{calafiore2011effect} of inactivity. Using shorter cutoff times to indicate inactivity would only reduce the recorded time spent browsing, and thus increase overestimation of browsing activity. In Figure \ref{fig:deltaold} in Appendix \ref{app.analysis}, we consider the impact of using 5 minutes of inactivity as a cutoff. Since this only reduces the recorded time spent browsing, unsurprisingly, we find that overestimation of browsing activity increases.

In the analysis above, we only counted time spent browsing on a laptop or desktop, as measured by our browsing extension, and did not include mobile browsing activity. Recent 2021 data \cite{broad} show that users spend 55.9\% of their browsing time on a desktop or laptop. We repeat the analysis above with this adjustment factor, by scaling down each participant's self-report by a factor of 0.559 to account for only measuring desktop/laptop browsing. Even after the adjustment, most participants still overestimate the amount of time they spend online, relative to our observational measurements (80.6\% of without adjustment vs. 77.4\% with adjustment). A more detailed analysis can be found in Appendix \ref{app.analysis}.

\begin{wraptable}{R}{0.5\textwidth}
\vspace{-0.5cm}
\small{
\begin{tabular}{|llc|}
\hline
\multicolumn{2}{|l}{Feature}                                                      & $p$-value                              \\ \hline
\multicolumn{2}{|l}{\textbf{Gender}}                                              &                                      \\
 & Male vs Female                                                                & 0.599                                  \\ \hline
\multicolumn{2}{|l}{\textbf{Race}}                                                &                                      \\
 & Asian vs Black or African American & 0.970                                  \\
 & Asian vs White                                                                & 0.420                                  \\
 & Black or African American vs White  & 0.511                                  \\ \hline
\multicolumn{2}{|l}{\textbf{Age}}                                                 &                                      \\ 
 & 18-24 vs 25-34                                                                & 0.433                                  \\
 & 18-24 vs 35-44                                                                & 0.791 \\
 & 25-34 vs 35-44                                                                & 0.659 \\ \hline
\end{tabular}
}
    \caption{$p$-values for pairwise $t$-test for equality of means for perceptions $\delta_i$s across demographic groups. 
    }
    \label{tab:table_pval_demovar_pair_ttest_delta}
\vspace{-1cm}
\end{wraptable}

\paragraph{Demographic variance.} 
We additionally investigate whether the biases in participants' perceptions differ across demographic groups. For each feature, we test for equality of means for $\delta_i$ across groups. We find no significant difference among genders ($t=-0.536$, $p=0.599$), age groups (pairwise, $t=-0.796$, $p=0.433$; $t=0.300$, $p=0.791$; $t=0.508$, $p=0.659$), and races (pairwise, $t=0.038$, $p=0.970$; $t=-0.831$, $p=0.420$; $t=-0.673$, $p=0.511$). A complete presentation of these results is given in Table \ref{tab:table_pval_demovar_pair_ttest_delta}.

\subsection{Perceptions of browsing activity by website category}\label{s.percat}

Next, we investigate whether participants had correct perceptions about the type of websites they browse most frequently. In the pre-study survey, we asked participants ``What are your most frequented categories of websites to visit?'' Each participant could select as many as they desired from the list of: Social Network, Business, Entertainment, News, Search, Banking, Shopping, Blogging, and Reference. These options correspond to categories on Alexa Top Websites by Category \cite{alexa} (see Section \ref{s.analysis} for details). Each participant $i$ reported their $k_i$ most frequently visited categories; we compared this with their top $k_i$ 
website categories, as measured by total time spent browsing. Participants on average chose 4.53 categories (SD = 1.34). See Table \ref{tab:oldcat_users_count} in Appendix \ref{app.figntab} for the number of participants who chose each category and the number of participants for whom each category was among their top $k_i$. %

Table \ref{tab:confusion_mtrx_alexa} presents a confusion matrix that summarizes whether participants' self-reported top browsing categories were among their actual most browsed categories.  Table \ref{tab:confusion_mtrx_alexa} also shows the percentage of participants with correct and incorrect perceptions for each category. 
 In Table \ref{tab:confusion_mtrx_alexa}, orange shaded cells indicate incorrect perceptions. Specifically, the orange shaded column on the left (observed in top $k_i$ categories of browsing, but not self-reported in top $k_i$) corresponds to participants who underestimated the amount of time they spent on each category, which indicates false positive rate. The orange shaded column on the right (not observed in top $k_i$ categories of browsing, but self-reported to be in top $k_i$) corresponds to participants who overestimated the time they spent on each category, which indicates false negative rate. 

We observe that while participants over-report their time online (see prior section), they are relatively accurate in their perceptions of where they spend their time. %
Out of 145 total top categories reported in total by our 31 participants, 50.3\% truly were top categories of the participant's observed behavior. The categories are sorted in Table \ref{tab:confusion_mtrx_alexa} by accuracy of participant perceptions. We see that participants had the most accurate perceptions of their browsing on Blogging and News websites, and the least accurate perceptions of Shopping and Business websites. We also see that most of the error in perceptions came from participants overestimating their level of browsing a particular category (i.e., false positive), which happened uniformly across website categories.

\begin{table*}[tbh]
\centering
\small
\begin{tabular}{|l|l|l|l|l|l|l|}
\hline
                           & \multicolumn{2}{l|}{\begin{tabular}[c]{@{}l@{}}{Observed Among Top $k_i$} \end{tabular}}                                & \multicolumn{2}{l|}{\begin{tabular}[c]{@{}l@{}}{Observed Not Among Top $k_i$} \end{tabular}}                                   & \cellcolor[HTML]{B4D4FF}                                                                               & \cellcolor[HTML]{FFCE93}                                                                             \\ \cline{2-5}
\multirow{-2}{*}{Category} & {\begin{tabular}[c]{@{}l@{}}Self-Report\\Top $k_i$\end{tabular}}  & {\begin{tabular}[c]{@{}l@{}}Self-Report\\Not Top $k_i$\end{tabular}} & {\begin{tabular}[c]{@{}l@{}}Self-Report\\ Top $k_i$\end{tabular}} & {\begin{tabular}[c]{@{}l@{}}Self-Report\\ Not Top $k_i$\end{tabular}} & \multirow{-2}{*}{\cellcolor[HTML]{B4D4FF}\begin{tabular}[c]{@{}l@{}}Correct\\ perception\end{tabular}} & \multirow{-2}{*}{\cellcolor[HTML]{FFCE93}\begin{tabular}[c]{@{}l@{}}Incorrect\\ perception\end{tabular}} \\ \hline
Blogging                    & \cellcolor[HTML]{B4D4FF}0\%        & \cellcolor[HTML]{FFCE93}3.1\%       & \cellcolor[HTML]{FFCE93}12.5\%       & \cellcolor[HTML]{B4D4FF}84.4\%      & 84.4\%                                                                                                   & 15.6\%                                                                                                 \\ \hline
News                       & \cellcolor[HTML]{B4D4FF}0\%        & \cellcolor[HTML]{FFCE93}3.1\%       & \cellcolor[HTML]{FFCE93}18.8\%       & \cellcolor[HTML]{B4D4FF}78.1\%      & 78.1\%                                                                                                   & 21.9\%                                                                                                \\ \hline
Search                     & \cellcolor[HTML]{B4D4FF}65.6\%     & \cellcolor[HTML]{FFCE93}9.4\%     & \cellcolor[HTML]{FFCE93}18.7\%       & \cellcolor[HTML]{B4D4FF}6.3\%     & 71.9\%                                                                                                 & 28.1\%                                                                                               \\ \hline
Social Network             & \cellcolor[HTML]{B4D4FF}53.1\%       & \cellcolor[HTML]{FFCE93}9.4\%       & \cellcolor[HTML]{FFCE93}18.7\%       & \cellcolor[HTML]{B4D4FF}18.8\%      & 71.9\%                                                                                                   & 28.1\%                                                                                                 \\ \hline
Banking                    & \cellcolor[HTML]{B4D4FF}6.3\%        & \cellcolor[HTML]{FFCE93}0\%       & \cellcolor[HTML]{FFCE93}31.2\%       & \cellcolor[HTML]{B4D4FF}62.5\%      & 68.8\%                                                                                                   & 31.2\%                                                                                                 \\ \hline
References                 & \cellcolor[HTML]{B4D4FF}28.1\%        & \cellcolor[HTML]{FFCE93}15.6\%       & \cellcolor[HTML]{FFCE93}21.9\%       & \cellcolor[HTML]{B4D4FF}34.4\%      & 62.5\%                                                                                                   & 37.5\%                                                                                                 \\ \hline
Entertainment              & \cellcolor[HTML]{B4D4FF}56.3\%       & \cellcolor[HTML]{FFCE93}9.4\%       & \cellcolor[HTML]{FFCE93}28.1\%       & \cellcolor[HTML]{B4D4FF}6.2\%      & 62.5\%                                                                                                   & 37.5\%                                                                                                 \\ \hline
Shopping                   & \cellcolor[HTML]{B4D4FF}12.5\%       & \cellcolor[HTML]{FFCE93}6.3\%      & \cellcolor[HTML]{FFCE93}40.6\%       & \cellcolor[HTML]{B4D4FF}40.6\%      & 53.1\%                                                                                                   & 46.9\%                                                                                                 \\ \hline
Business                   & \cellcolor[HTML]{B4D4FF}6.3\%       & \cellcolor[HTML]{FFCE93}18.7\%      & \cellcolor[HTML]{FFCE93}34.4\%        & \cellcolor[HTML]{B4D4FF}40.6\%      & 46.9\%                                                                                                   & 53.1\%                                                                                                 \\ \hline
\end{tabular}
\caption{Confusion matrix showing accuracy of participants perceptions regarding the website categories they most frequently browse. Each participant $i$ self-reported their $k_i$ top website categories, and these were compared with their top $k_i$ categories of observed browsing based on time spent browsing. Blue shaded cells indicate correct perceptions (true positives or true negatives), and orange shaded cells indicate incorrect perceptions (false positives or false negatives). Total correct and incorrect perceptions are also calculated for each category. }

\label{tab:confusion_mtrx_alexa}
\end{table*}

%% file: rq3.tex
\section{RQ3: Do people change browsing behavior if they are aware of being observed?}

In this section, we aim to test whether people behave differently online when they are consciously aware of being observed. We hypothesize that participants' conscious awareness of being observed may be the most salient early in the study, shortly after they provide informed consent for data collection, and that this awareness may diminish over time. 
If this were the case, we would expect to observe a change in participants' behavior over time during the study. As a proxy measure for this analysis, we test a hypothesis that the distribution of participants' activity during the first half of the study (Days 1-7) is different from the second half (Days 8-14). We first test for changes in level of browsing activity in Section \ref{s.obsact}, and then for differences in website categories browsed in Section \ref{s.obscat}.

\subsection{Changes in level of activity}\label{s.obsact}

\begin{figure}[tbh]
	\centering
	\subfloat[][Number of browsing actions]{\includegraphics[width=.35\textwidth]{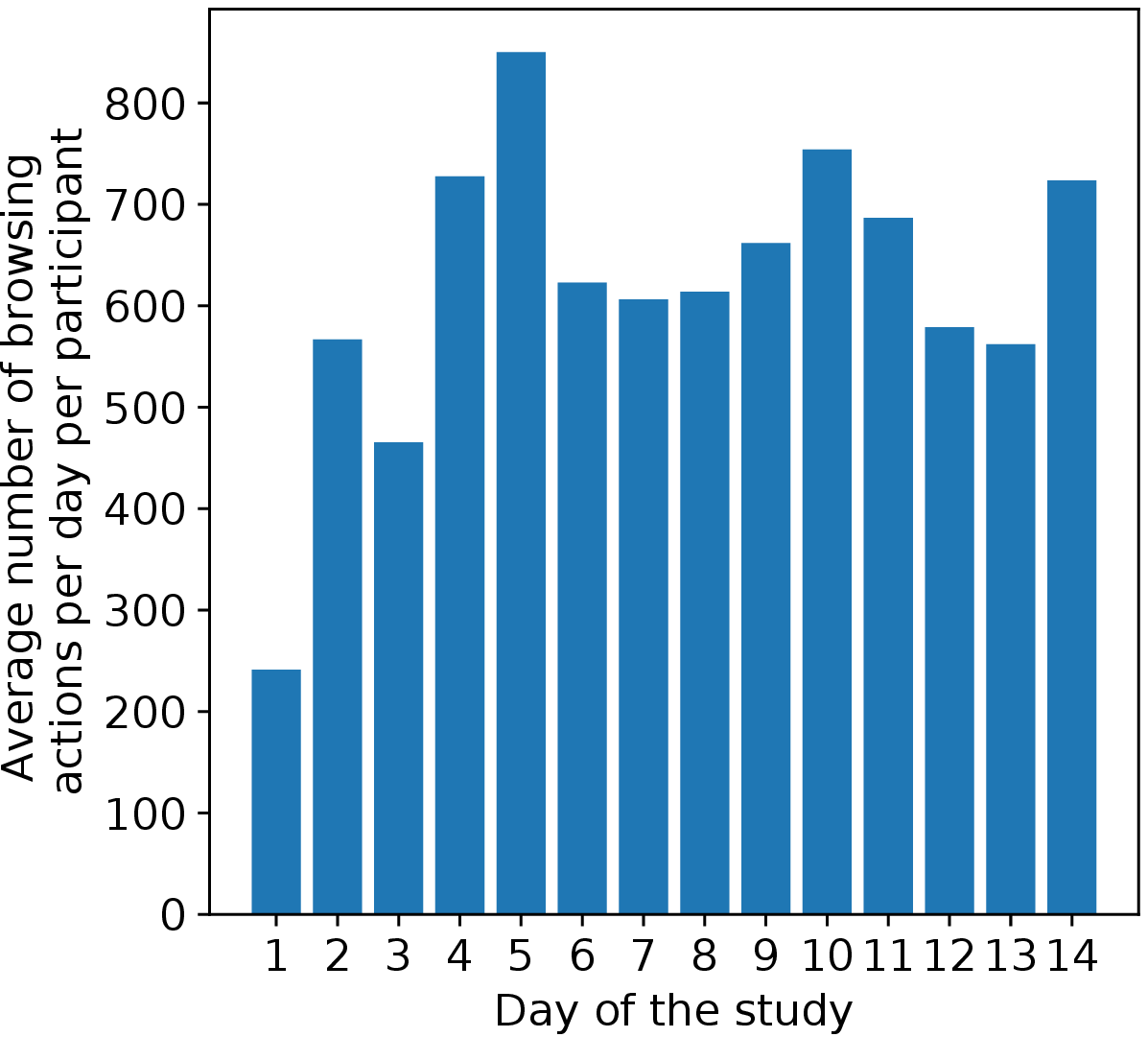}}
	\subfloat[][Time spent browsing]{\includegraphics[width=.35\textwidth]{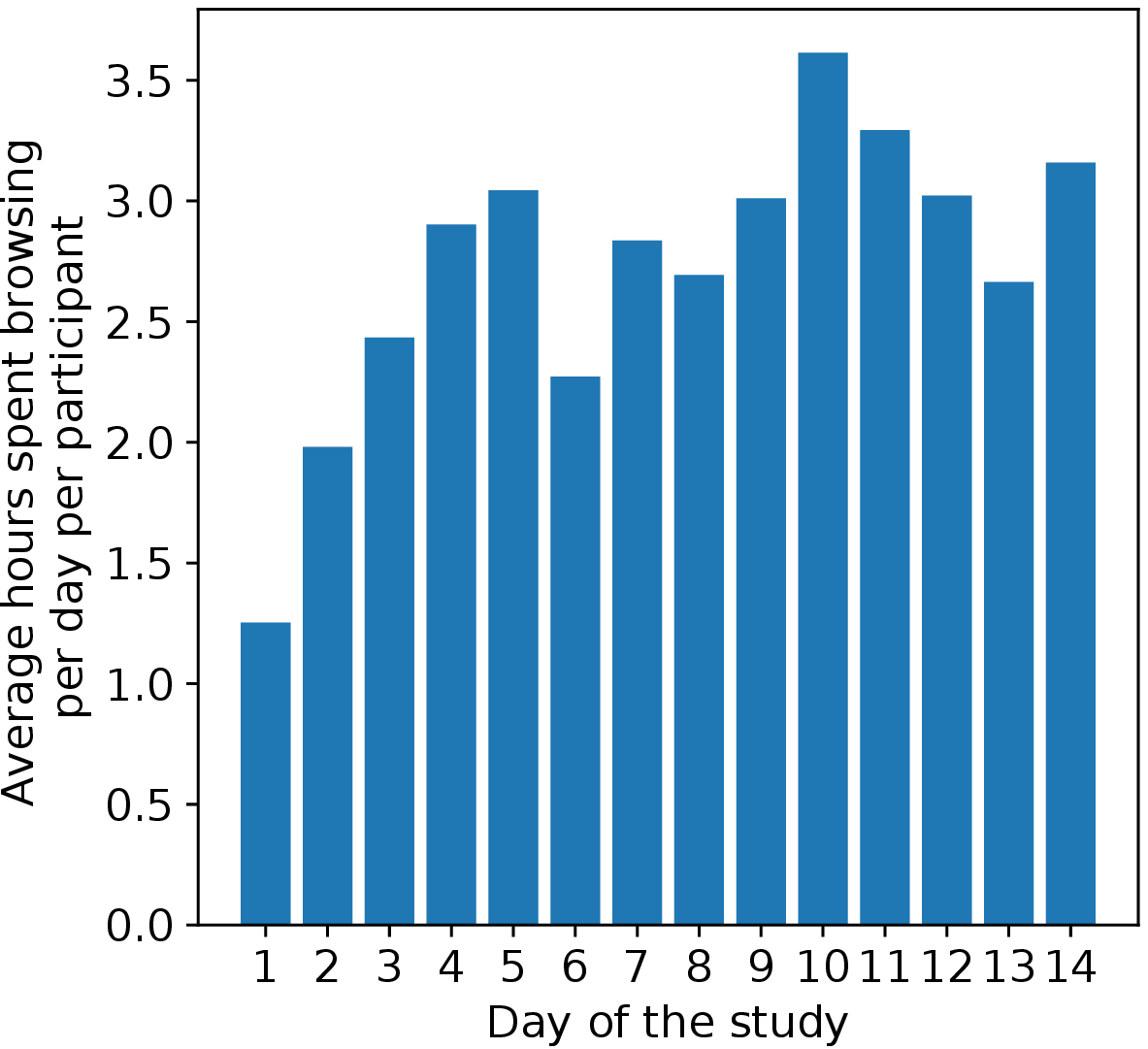}}	
	\caption{Average number of actions and time spent browsing, per participant per active browsing day of the study.
	} \label{fig:actions_pday}
\end{figure}

We investigate whether participants changed their level of browsing activity during the course of the study. We use as activity metrics both daily average number of browsing actions and daily average browsing time. Figure \ref{fig:actions_pday} shows the daily average number of browsing actions and the daily average browsing time, both averaged across all participants. We note that participants installed the browsing extension during the first day of the study, so browsing activity is noticeably lower on Day 1 since a full day of browsing was not captured. %

While we observe variance in average daily activity, we do not find significant differences in the level of browsing activity observed over the duration of the study. Specifically, under the activity metric of daily average number of browsing actions per active browsing day, we find that neither the mean number of actions ($t=-0.915$, $p=0.348$) nor variance in number of actions ($L=2.009$, $p=0.182$) differs significantly between the first and second half of the study. Under the metric of daily average browsing time, similarly, both mean of the number of hours ($-1.230$, $p=0.208$) and variance  ($L=2.191$, $p=0.165$) do not show a significant difference between the first and second half of the study. Participants' reports in our post-study survey support these findings, with only 2 of 31 participants reporting that that they altered their behavior during the study. 

\subsection{Changes in type of web use}\label{s.obscat}
\begin{figure}[H]
	\centering
	\subfloat[][Browsing actions]{\includegraphics[width=.49\columnwidth]{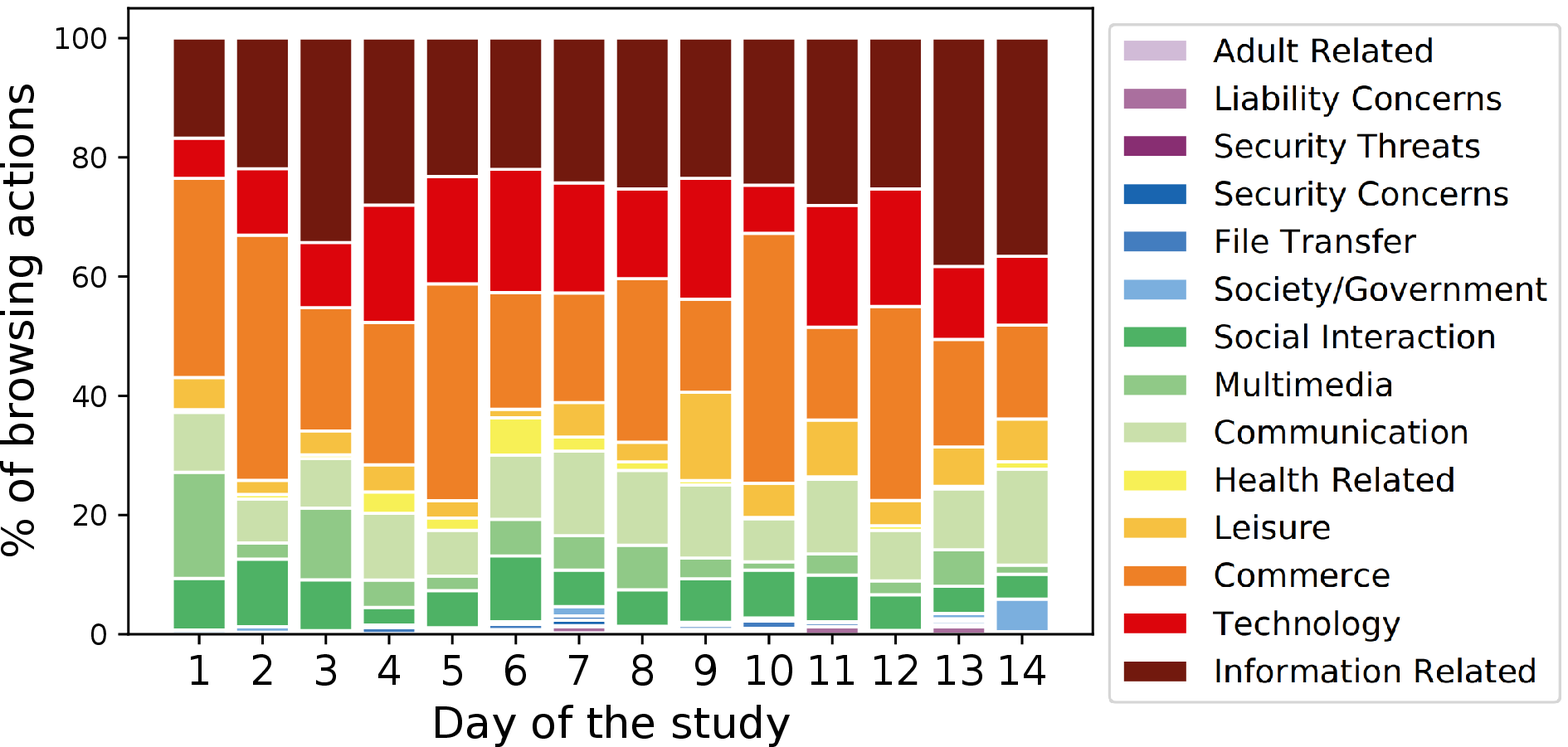}}
	\subfloat[][Time spent browsing]{\includegraphics[width=.49\columnwidth]{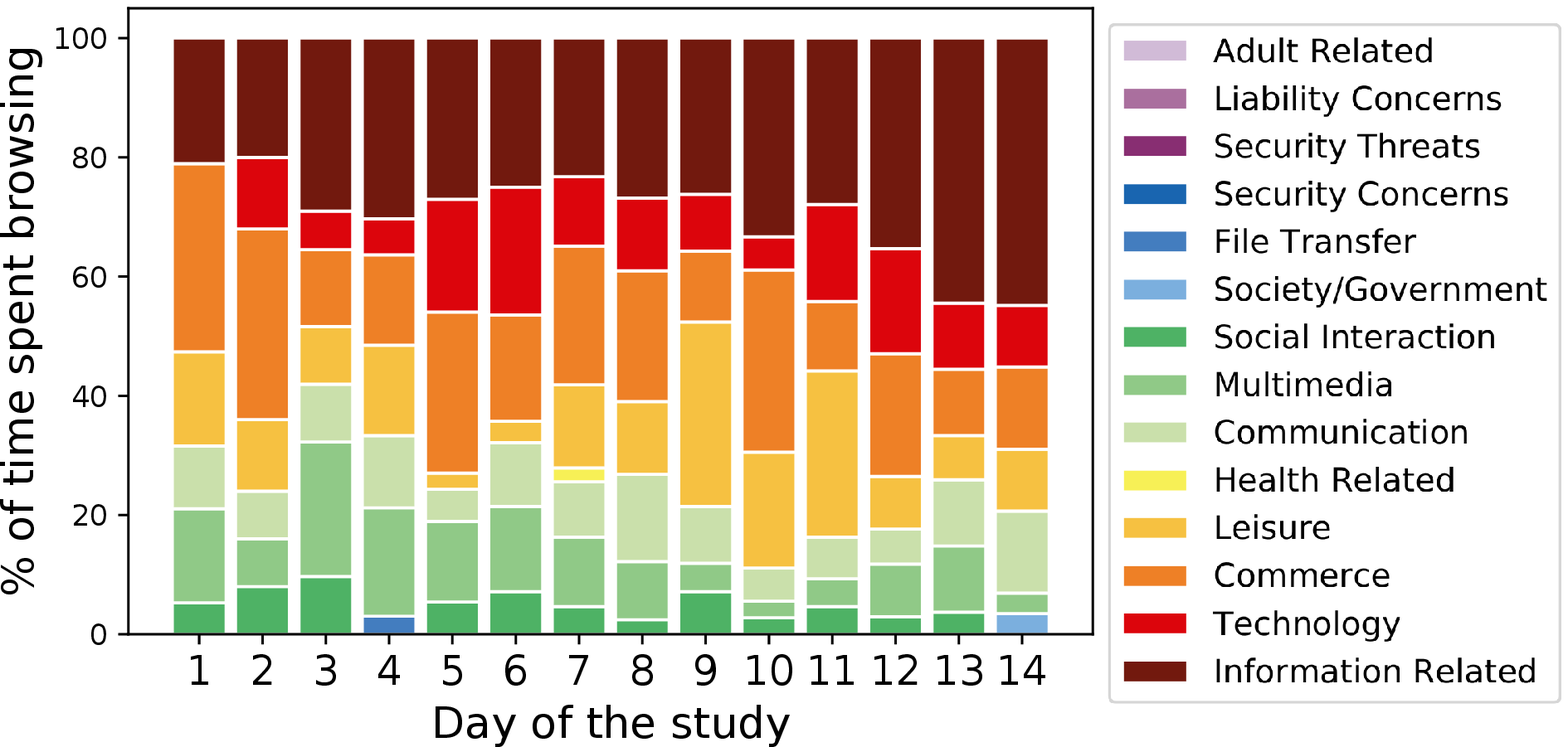}}
	\\
\caption{Proportion of (a) browsing actions and (b) time spent browsing on each website category on each day of the study. 
} \label{fig: cat_all_gen}
\end{figure}

We also investigate whether participants' browsing activity across website categories changed over the course of the study. 
Figure \ref{fig: cat_all_gen} shows the proportion of browsing activity across website categories as measured by both the number of browsing actions and hours spent browsing. 
With number of browsing actions, we do not observe a significant change in distribution of web use across website categories between the first half of the study (Days 1-7) and the second half (Days 8-14) ($W=37.0$, $p=0.357$); similarly, under time spent browsing, we do not observe a significant difference ($W=38.0$, $p=0.390$).

%% file: discussion.tex
\section{Discussion and Conclusions}\label{s.discussion}
In this work we provide an up-to-date picture of a young (under 45 years old) sample of internet users' browsing behavior (RQ1). We find that people are viewing similar numbers of pages today as in prior work but are spending significantly more time online (those in our sample spent an average of three hours a day online compared to prior work conducted nearly a decade ago, which observed an average of one hour of daily online activity~\cite{abramson2013associative,kumar2010characterization}). Echoing prior work conducted nearly a decade ago~\cite{goel2012does}, we find relatively little demographic variance in browsing activity, although differently from prior work that leveraged demographic inference data~\cite{goel2012does}, we do observe that the older users (35-44) in our sample browse significantly less than those who are younger.

Our work adds to the body of knowledge on digital browsing behavior in that we examine not only how much time people spend online and how many pages they view, but \textit{what} they do online. Prior work has studied user behavior in terms of webpage access time \cite{abramson2013associative} and number of page revisits \cite{kumar2010characterization}. To the best of our knowledge, this is the first work to study user behavior in terms of the proportion of types of actions performed on websites and across different website categories.
We find that people spend most of their time on Information Related, Commerce, Technology, Leisure, and Communication websites, with Social Interaction (social media) websites ranking seventh. While people spend different amounts of time on different types of pages, they behave quite similarly in terms of the actions they take (clicks, typing, urlChanges) on these pages. There are a few exceptions: Multimedia websites see less typing and urlChanges, as people are primarily clicking on and watching videos; File Transfer websites see a high number of urlChanges characterizing file uploads; and Security Concern websites, which include suspected phishing URLs and misspellings of popular URLs, and can be characterized by a high number of urlChanges, in line with prior findings that malicious websites aim to redirect users to additional malicious pages and opportunities for credential capture~\cite{thomas2011design}. These findings suggest that one potentially promising direction for augmenting existing approaches~\cite{cao2014behavioral, cao2015detecting,dong2008user} to keeping people safe online is to add common website interaction patterns as signals for detecting malicious websites.

Further, our work addresses a critical question for the study of online behavior: we examine the relationship between participants' self-reported online browsing --- in terms of time spent online and types of web uses --- and their actual behavior as we observe it using our measurement tools (RQ2). We find that participants significantly over-report their daily time spent online, by an average of 4.5 hours per day; this overreporting does not vary with demographics (age, gender, or race). This finding aligns with prior work that examined the accuracy of people's self-reports about their Facebook behavior, specifically, finding that people overestimated their time spent on the platform~\cite{10.1145/3313831.3376435}. This suggests that findings regarding the relationship between various digital constructs (e.g., social capital, digital skill~\cite{hargittai2010digital, ellison2007benefits}) and self-reported time spent online should be interpreted with care: people's perceptions of how much time they spend online may over-represent the time they actually spend online.

While participants in our study over-reported their time spent online, they were relatively accurate in their reports about the types of websites where they spent the most time. This suggests, in line with prior work examining the accuracy of people's self reports about the speed with which they update their computers~\cite{redmiles2018asking}, that people may have an accurate \textit{relative} sense of their digital behavior, but inaccurate absolute perceptions (i.e., about the exact amount of time they spend online or the precise strength of their passwords~\cite{wash2017can}). This suggests that observational methods of measurement may be most appropriate for use when precise absolute measurements are necessary, but that self-report measurements may be an appropriate proxy when only relative measurements are required.

Finally, given prior findings from other fields on possible observational biases that may occur when participants are aware that their behavior is being observed~\cite{AB09}, we examine whether participants' observed behavior changed over the course of our experiment to see whether we could detect such observational biases in our measurements of web behavior (RQ3). We find no significant changes in participant behavior over the course of the study. It is possible that we observe no behavior change because 14 days is not a sufficiently long period of time for participants to forget that they are being observed. Alternately, people may have such a pervasive sense of being observed online~\cite{duffy2019you} that even installing a browser plugin that they know observes their behavior may not change their activity. Future work is necessary to further explore the question of observation bias in measurements of digital behavior, perhaps through comparison of proprietary industry measurement data -- which a user is not actively aware is being collected -- with measurement data from a disclosed browser plugin such as the one we use in this study.

\paragraph{Acknowledgements.} The authors wish to thank Elizabeth Krizay for her contributions to designing the study procedures.

%% file: appendix.tex
\section{Additional figures and tables} \label{app.figntab}

\begin{figure}[tbh]
	\centering
	{\includegraphics[width=0.6\textwidth]{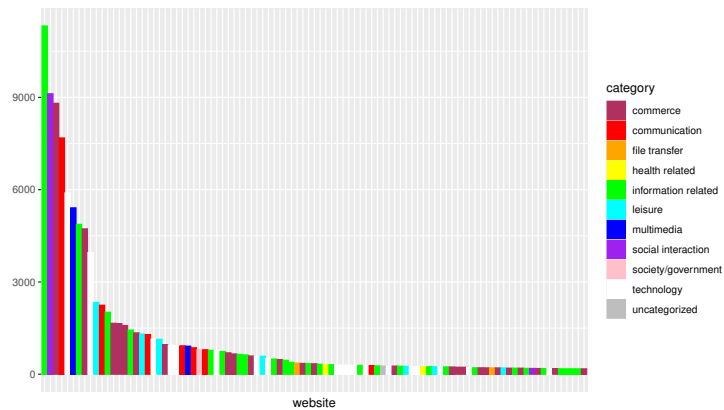}}
\caption{Distribution of browsing actions performed on the 100 most browsed websites in the study (as measured by number of browsing actions), color-coded by category. 
} \label{fig:old_hist_url}
\end{figure}

Figure \ref{fig:old_hist_url} shows the distribution of browsing actions and website category for the top 100 most browsed websites (as measured by number of browsing actions) by all users during the study. Each website is color-coded to indicate the category of that website. We observe that the top nine websites have high levels of activity, and that the level of activity drops off quickly in the distribution to leave a long tail.

\begin{table}[tbh]
\centering
\include{table_p-values_corrected}
\caption{$p$-values for Pearson's $\chi^2$-test for homogeneity based on distribution of browsing actions within websites. Tests were performed pairwise for all website categories.}
\label{tab:p-values}
\end{table}

Table \ref{tab:p-values} supports the analysis of RQ1 in Section \ref{s.categories}, and presents the $p$-values of pairwise tests for differences in browsing behavior across website categories. Specifically, it shows the $p$-values for Pearson's $\chi^2$-test for homogeneity of the distribution of browsing actions within each category, as illustrated in Figure \ref{fig: actions_inside_site}, across all pairs of website categories.

\renewcommand{\thesubfigure}{\arabic{subfigure}}
\begin{figure}[H]
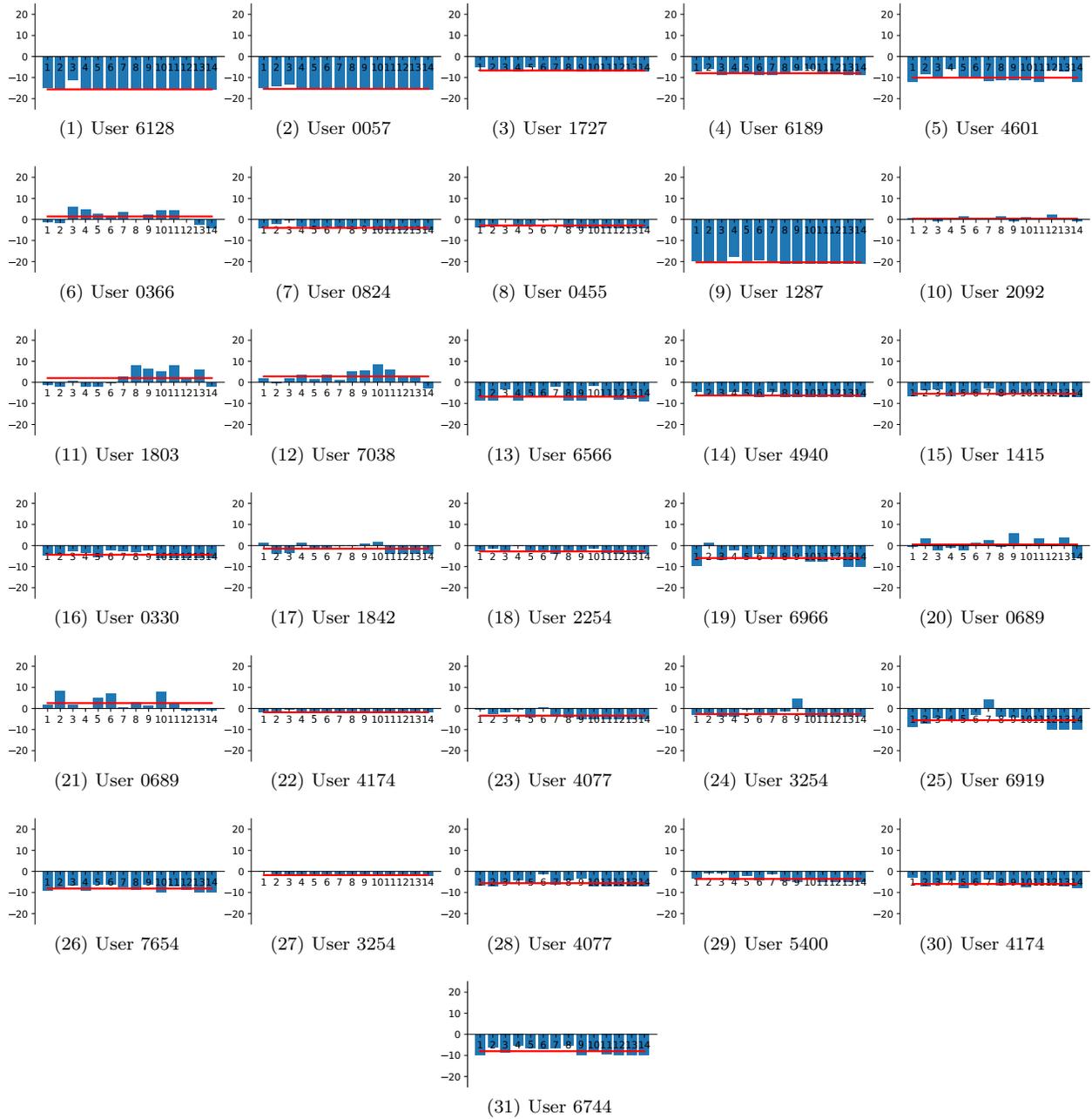

	\centering
	\include{figure_delta_hours_pday_puser}
	\caption{Differences between the actual number of hours spent browsing per day and self-reported number of browsing hours per day, for each participant in the study. On non-active browsing days, the time spent browsing was set to zero. The $x$-axis enumerates the day of the experiment. The red horizontal line is a mean of these differences over the days of experiment.} \label{fig:delta_hours_pday_puser}
\end{figure}

\renewcommand{\thesubfigure}{\alph{subfigure}}

Figure \ref{fig:delta_hours_pday_puser} supports the analysis of RQ2 in Section \ref{s.pertime} by providing a more detailed visualization of the $\delta_i$s in Figure \ref{fig:delta}a at a per-participant level. Each subfigure corresponds to a single participant, with their $\delta_i$ shown for each day of the study. Recall that $\delta_i=0$ corresponds to perfectly accurate perceptions of time spent browsing, $\delta_i<0$ (resp. $\delta_i>0$) corresponds to an overestimation (resp. underestimation) of browsing time. The red line in each subfigure illustrates the participant's average $\delta_i$ across all days in the study. We see that most users overestimate their time spent browsing, some by small amounts and some by large amounts.

Table \ref{tab:oldcat_users_count} supports the analysis of RQ2 in Section \ref{s.percat} by showing the number of participants who selected each category in the pre-study survey as one of their most browsed categories, and the number of participants who were observed to have each website category as one of their most browsed categories. Recall that in the pre-study survey, participants could select as many categories as they wished.  For participant $i$ who selected $k_i$ categories in the pre-study survey, we included their $k_i$ most browsed categories in the latter evaluation.

\begin{table}[tbh]
\centering
\begin{tabular}{|c|c|c|}
\hline
\textbf{Website category} & \textbf{\begin{tabular}[c]{@{}c@{}}Number of participants who \\ chose category as among\\ $k_i$ most frequently browsed\end{tabular}} & \textbf{\begin{tabular}[c]{@{}c@{}}Number of participants \\with category in top $k_i$\\ of actual browsing\end{tabular}} \\ \hline
Shopping          & 17                                                                                                   & 6                                                                                                                              \\ \hline
Reference         & 16                                                                                                   & 14                                                                                                                             \\ \hline
Social Network    & 23                                                                                                   & 20                                                                                                                             \\ \hline
Entertainment     & 27                                                                                                   & 21                                                                                                                             \\ \hline
Business          & 13                                                                                                   & 8                                                                                                                              \\ \hline
Search            & 27                                                                                                   & 24                                                                                                                             \\ \hline
News              & 6                                                                                                    & 1                                                                                                                              \\ \hline
Banking           & 12                                                                                                   & 2                                                                                                                              \\ \hline
Blogging          & 4                                                                                                    & 1                                                                                                                              \\ \hline
\end{tabular}
\caption{Alexa Top Websites \cite{alexa} categories offered in the pre-study survey, along with number of participants who named each category as among their most frequently browsed and number of participants for whom each category was among their observed top categories of browsing during the study.}
\label{tab:oldcat_users_count}
\end{table}

\section{Alternative Methodologies for RQ2}\label{app.analysis}

In this section, we consider two alternative methodologies for measuring the difference between participants' perceived and actual time spent browsing.

\paragraph{Using 5 minutes of inactivity as a cutoff.} We first consider using 5 minutes of inactivity as a cutoff to end an active browsing session, rather than 30 minutes as in Section \ref{s.pertime}. Intuitively, this will shorten each browsing session by 25 minutes, as participants will be considered inactive sooner after their last browsing action. This alternative methodology gives a more accurate measure of browsing activities that involve the actions listed in Table \ref{tab:action-types-class}, but may be less likely to capture passive browsing experiences, such as watching a video or reading a long article.

\begin{figure}[tbh]
	\centering
	\subfloat[][\small Observed vs perceived browsing time]{\includegraphics[width=.45\textwidth]{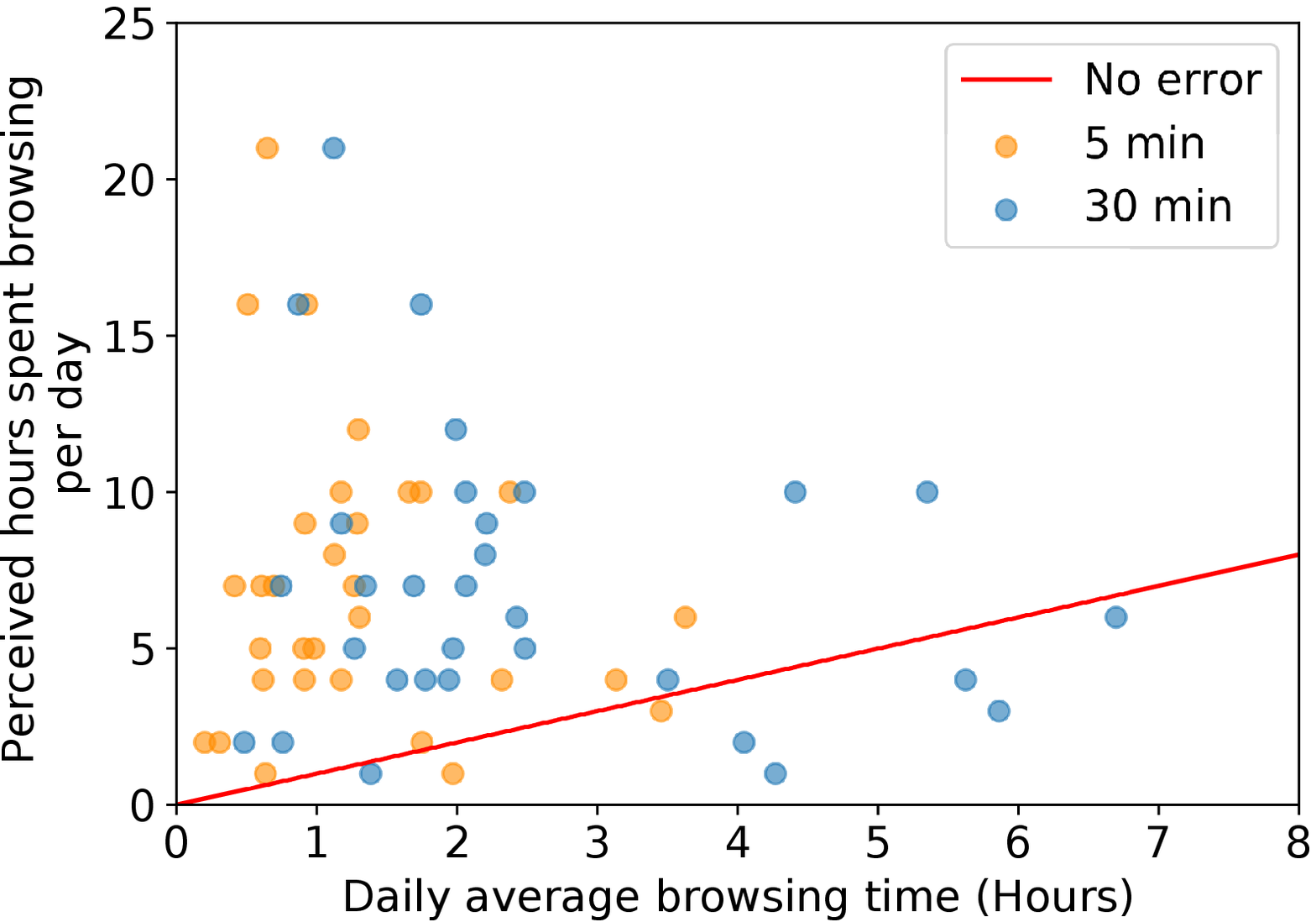}}
	\subfloat[][\small Distribution of error in perceptions]{\includegraphics[width=.45\textwidth]{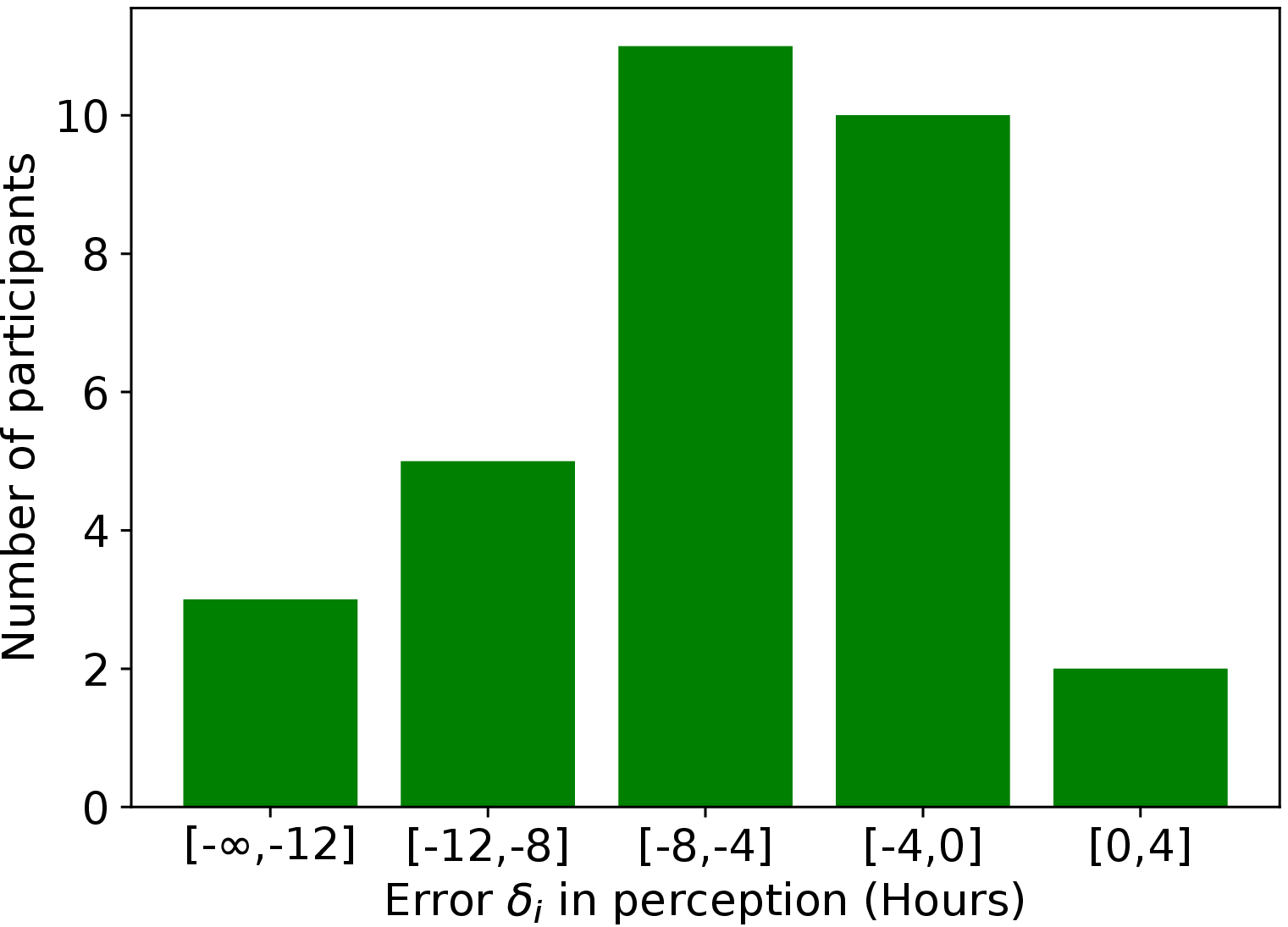}}
\caption{
(a) Scatter plot illustrating observed daily average browsing time vs. perceived (self-reported) number of hours spent browsing per day. Each point corresponds to one participant. Orange dots correspond to analysis with 5 minutes of inactivity as a cutoff, and the blue dots correspond to analysis with 30 minutes as a cutoff time as in Section \ref{s.pertime}. Red line $x=y$ corresponds to no error in perceptions. 
(b) Distribution of error values $\delta_i$ in the participant population using 5 minutes of inactivity as a cutoff. 
} \label{fig:deltaold}
\end{figure}

Similar to the findings in Section \ref{s.pertime}, we  find  that  the  majority  of  participants  (29  out  of  31,  93.55\%)  overestimate their daily browsing time.  Figure \ref{fig:deltaold} is analogous to Figure \ref{fig:delta}, as it visualizes the relationship between participants' actual time spent browsing and their perceived time spent browsing. Figure \ref{fig:deltaold}a shows a scatter plot of the observed daily average browsing time versus the perceived (self-reported) number of hours spent browsing per day, with one point corresponding to each participant. For ease of comparison with the results of Section \ref{s.pertime}, the orange dots correspond to analysis with 5 minutes as a cutoff time, and the blue dots correspond to analysis with 30 minutes as a cutoff time. The red line $x=y$ corresponds to perfectly accurate perceptions.  Figure \ref{fig:deltaold}b shows the distribution of error $\delta_i$ in hours among participants. Recall that $\delta_i$ is the difference between participant $i$'s observed daily average browsing time (now, as measured using a 5 minute cutoff for inactivity) and the number of hours per day they reported to spend browsing in the pre-study survey. Using a $t$-test, we find that the mean of the $\delta_i$s among participants is significantly different from 0 ($t=-6.497$, $p<10^{-7}$), which implies that participants still do not have accurate perceptions of their active browsing time, even under this alternative analysis method.

We additionally investigate whether the biases in participants’ perceptions differ among demographic groups.  We find no significant difference between $\delta_i$s for different genders ($t=-0.595$, $p=0.505$),  age groups ($t=-.724$, $p= 0.487$; $t=0.220$, $p= 0.607$; $t=0.394$, $p= 0.531$),  and races ($t=0.223$, $p =0.794$; $t=-0.803$, $p =0.537$; $t=-0.794$, $p =0.428$). These results are presented in Table \ref{tab:table_pval_demo_pair_ttest_actions_adj}.

\paragraph{Adjustments for desktop versus mobile browsing.} Our pre-study survey asked participants about their perceived time spent browsing, without distinguishing between desktop\footnote{Desktop here refers to devices that default to desktop versions of websites, which includes desktop and laptop personal computers, but does not include phones or tablets.} and mobile browsing, but our extension was only able to capture browsing on a desktop or laptop. Recent 2021 data \cite{broad} found that 55.9\% of users' browsing time is spent on a desktop device. We account for discrepancy by scaling down each participant's self-reported time spent browsing by a factor of 0.559 and repeating the analysis of Section \ref{s.pertime}.

\begin{figure}[H]
	\centering
    \subfloat[][\small Observed vs perceived browsing time]{\includegraphics[width=.45\columnwidth]{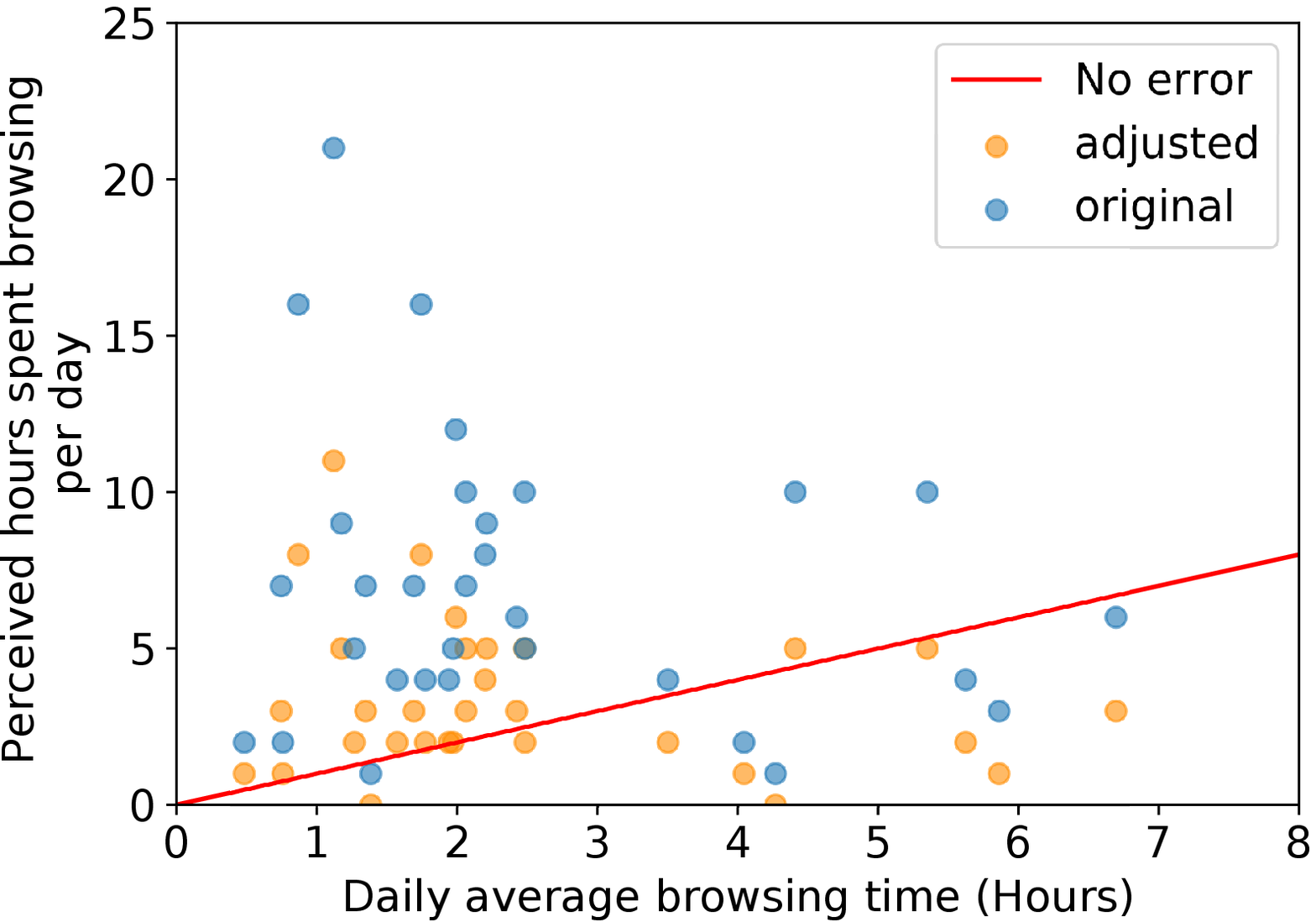}}
    \subfloat[][\small Distribution of error in perceptions]{\includegraphics[width=.45\columnwidth]{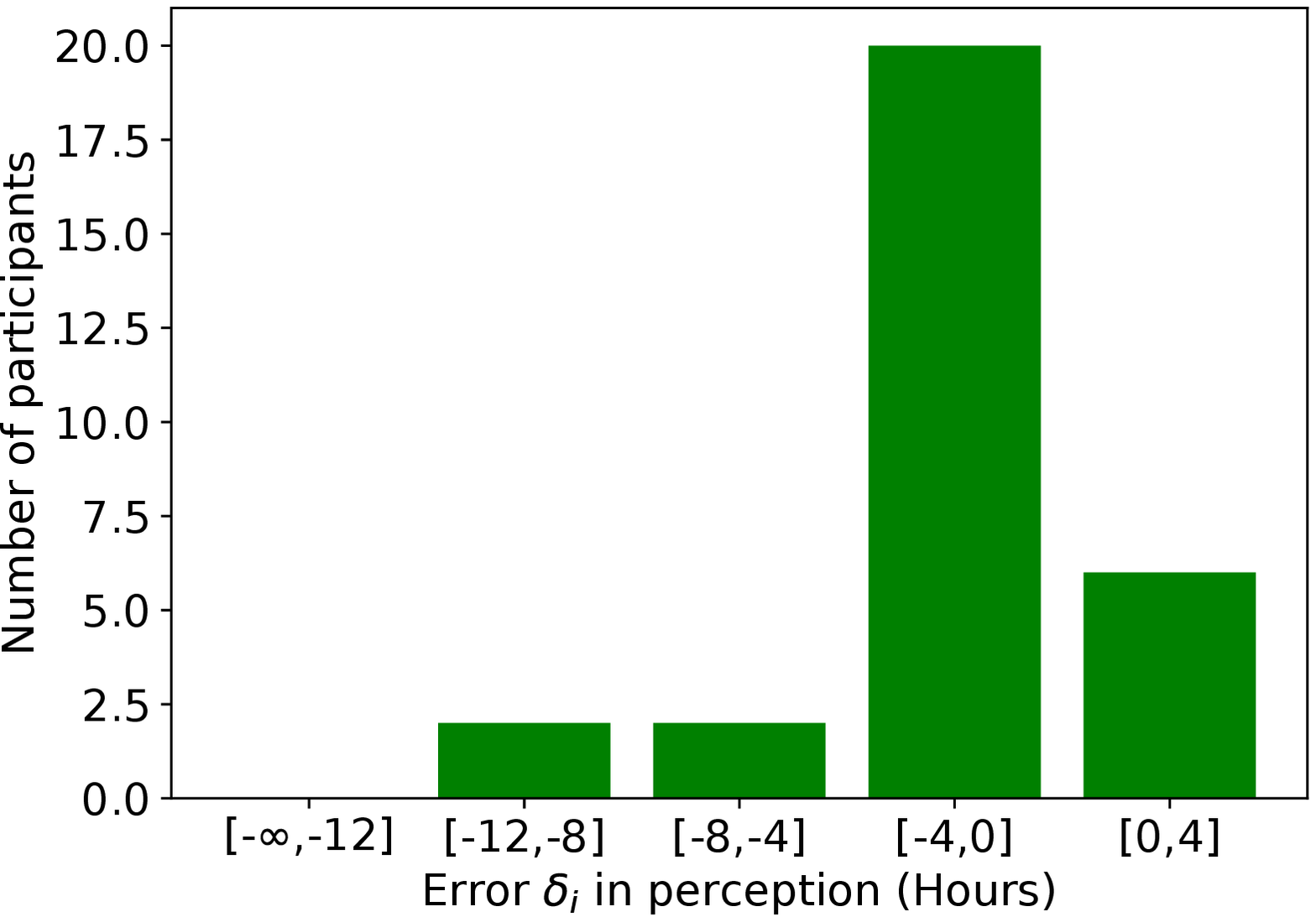}}
\caption{
(a) Scatter plot illustrating observed daily average browsing time vs. adjusted perceived (self-reported) number of hours spent browsing per day. Each point corresponds to one participant. Adjusted and original (non-adjusted, as in Section \ref{s.pertime}) points are shown. Red line $x=y$ corresponds to no error in perceptions. 
(b) Distribution of error values $\delta_i$ in the participant population based on the adjusted perceived values. The average error $\delta_i$ is -1.41 hours (SD=3.35), with 77.4\% of participants over-estimating their time spent online. 
} \label{fig:delta_adj}
\end{figure}

Even after the adjustment, most participants still overestimate the amount of time they spend online, relative to our observational measurements (80.6\% of without adjustment vs. 77.4\% with adjustment). However, the mean of adjusted error $\delta_i$s is lower (-1.41 hours) than for non-adjusted values (-4.5 hours), suggesting that while participants still overestimate their time spent browsing, they overestimate by a smaller amount, relative to no adjustment. Figure \ref{fig:delta_adj} is analogous to Figure \ref{fig:delta}, showing (a) a scatter plot of participants' observed daily average browsing time versus their adjusted perceived (self-reported) hours of daily browsing, and (b) distribution of errors $\delta_i$.

Using a $t$-test, we find that the mean of the adjusted $\delta_i$s is significantly different from 0 ($t=-2.348$, $p=0.026$). When we look for differences in perception errors across demographic groups, we find no significant difference between adjusted $\delta_i$s for different genders ($t=-0.485$, $p=0.621$), age groups ($t=-0.783$, $p= 0.446$; $t=0.465$, $p= 0.535$; $t=0.708$, $p= 0.472$), and races ($t=-0.125$, $p =0.937$; $t=-0.838$, $p= 0.479$; $t=-0.561$, $p= 0.569$). A complete presentation of these results is given in Table \ref{tab:table_pval_demovar_pair_ttest_delta_adj}.

\begin{table}[tbh]
    \begin{minipage}{.45\linewidth}

\small{
\begin{tabular}{|llc|}
\hline
\multicolumn{2}{|l}{Feature}                                                      & $p$-value                              \\ \hline
\multicolumn{2}{|l}{\textbf{Gender}}                                              &                                      \\
 & Male vs Female                                                                & 0.505                                  \\ \hline
\multicolumn{2}{|l}{\textbf{Race}}                                                &                                      \\
 & Asian vs Black or African American & 0.794                                  \\
 & Asian vs White                                                                & 0.537                                  \\
 & Black or African American vs White  & 0.428                                  \\ \hline
\multicolumn{2}{|l}{\textbf{Age}}                                                 &                                      \\ 
 & 18-24 vs 25-34                                                                & 0.487                                  \\
 & 18-24 vs 35-44                                                                & 0.607 \\
 & 25-34 vs 35-44                                                                & 0.531 \\ \hline
\end{tabular}}
      \caption{$p$-values for pairwise $t$-test for equality of means of perception errors $\delta_i$s across demographic groups using 5 minutes of inactivity as a cutoff.}
      \label{tab:table_pval_demo_pair_ttest_actions_adj}
    \end{minipage}%
    \hfill
    \begin{minipage}{.45\linewidth}
\small{
\begin{tabular}{|llc|}
\hline
\multicolumn{2}{|l}{Feature}                                                      & $p$-value                              \\ \hline
\multicolumn{2}{|l}{\textbf{Gender}}                                              &                                      \\
 & Male vs Female                                                                & 0.621                                  \\ \hline
\multicolumn{2}{|l}{\textbf{Race}}                                                &                                      \\
 & Asian vs Black or African American & 0.937                                  \\
 & Asian vs white                                                                & 0.479                                  \\
 & Black or African American vs white  & 0.569                                  \\ \hline
\multicolumn{2}{|l}{\textbf{Age}}                                                 &                                      \\ 
 & 18-24 vs 25-34                                                                & 0.446                                  \\
 & 18-24 vs 35-44                                                                & 0.535 \\
 & 25-34 vs 35-44                                                                & 0.472 \\ \hline
\end{tabular}
}
    \caption{$p$-values for pairwise $t$-test for equality of means of perception errors $\delta_i$s across demographic groups using adjusted self-reports of browsing activity. 
    }
    \label{tab:table_pval_demovar_pair_ttest_delta_adj}
    \end{minipage} 
\end{table}

\section{Screenshots of Study Materials}\label{app.screenshots}

\begin{figure}[tbh]
	\centering
	{\includegraphics[width=0.4\textwidth]{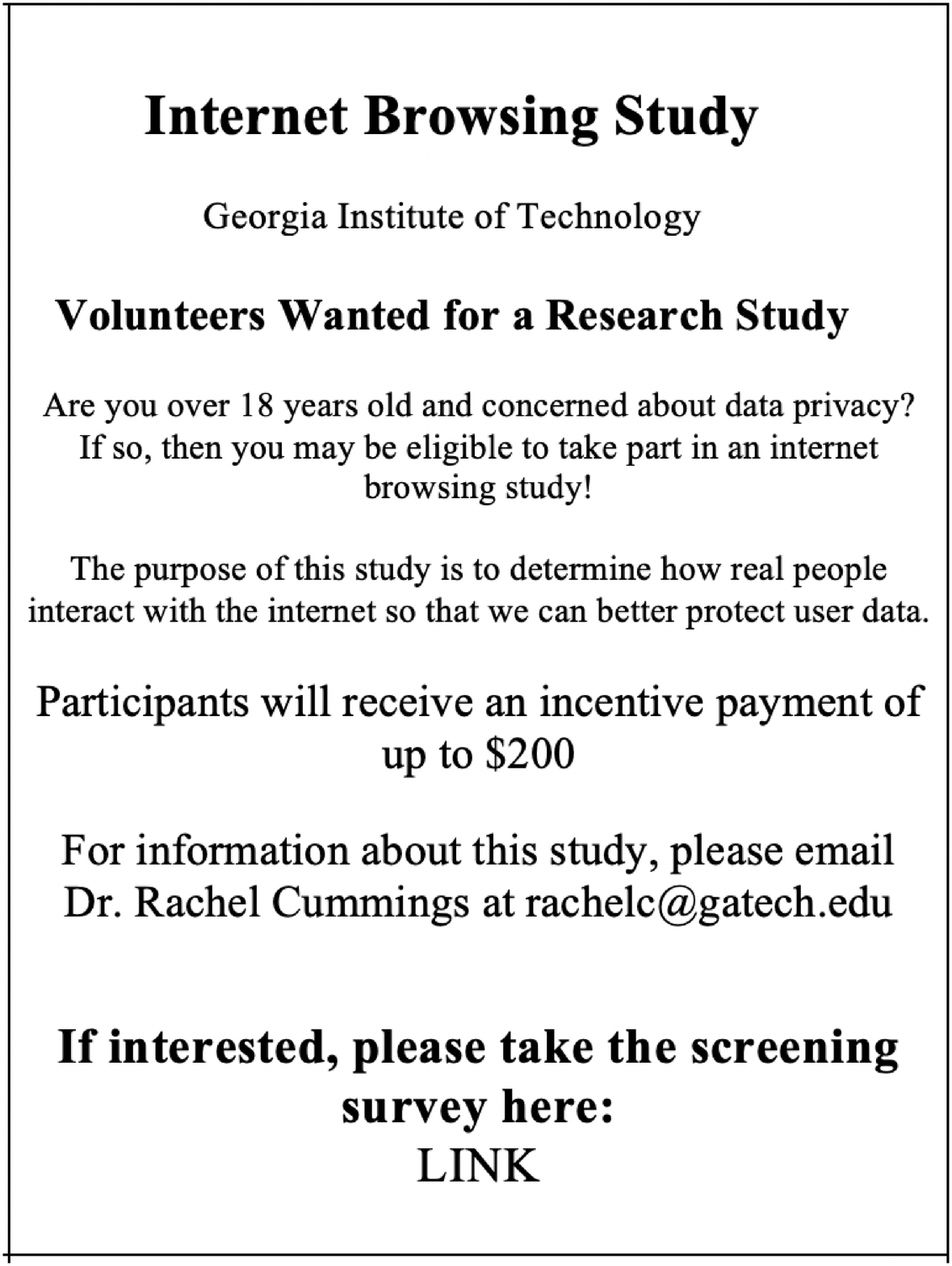}}
\caption{Recruitment flyer } \label{fig:flyer}
\end{figure}

\begin{figure}[H]
	\centering
    \subfloat[][\small Extension logo]{\includegraphics[width=.2\columnwidth]{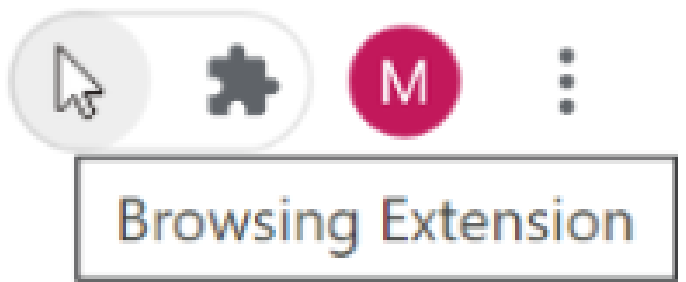}}
    \hspace{1em}
    \subfloat[][\small Extension menu displayed when logo is clicked]{\includegraphics[width=.25\columnwidth]{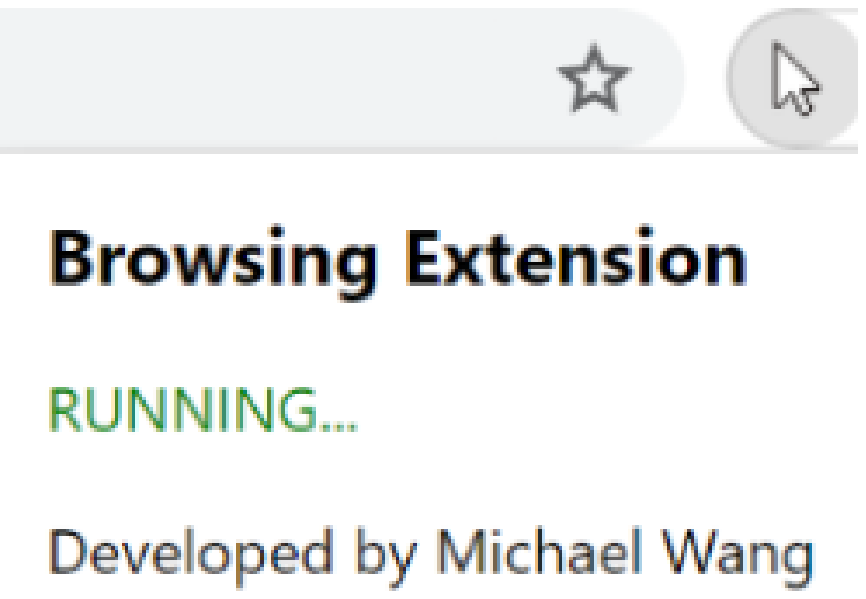}}
    \hspace{1em}
      \subfloat[][\small Extension information available in the Chrome Extensions page]{\includegraphics[width=.35\columnwidth]{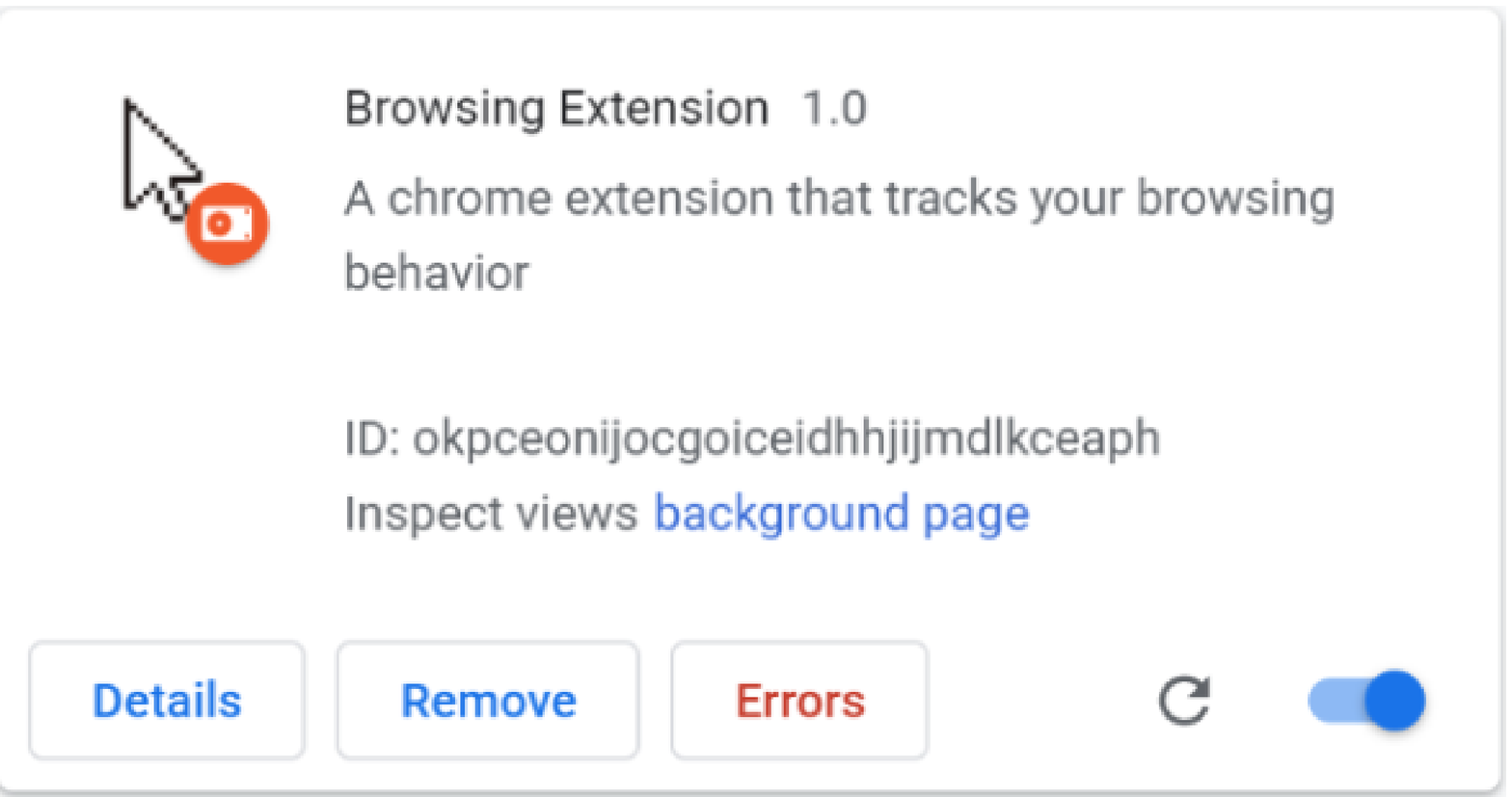}}
\caption{Screenshots of the browsing extension in the Chrome browser as seen by the participants during the study.
} \label{fig:browser}
\end{figure}

In this appendix, we show images related to participants' experience during the study. Figure \ref{fig:flyer} shows the recruitment flyer advertising the study that was used to recruit participants.  Figure \ref{fig:browser}(a) shows the extension logo that appeared continuously in the Chrome browser to the participants during the study, Figure \ref{fig:browser}(b) shows the extension menu that would appear if the participant clicked on the extension logo, and Figure \ref{fig:browser}(c) shows extension information that was viewable on the Chrome Extensions page. These were all designed to look generic and to neither reveal the purpose of the study, nor to remind participants that their browsing behaviors were being collected, to better address RQ3. This drove our design of the logo as simply a mouse cursor and the extension name as simply ``Browsing Extension''. Since the extension was designed to collect browsing data in the background without interfering with participant browsing, this was the only visual that participants experienced during data collection.

%% file: table_p-values_corrected.tex
\begin{adjustbox}{width=1.1\textwidth}
\begin{tabular}{|l|l|l|l|l|l|l|l|l|l|l|l|l|l|l|}
\hline
                                                                & \begin{tabular}[c]{@{}l@{}}Techno-\\ logy\end{tabular} & \begin{tabular}[c]{@{}l@{}}Information\\  Related\end{tabular} & \begin{tabular}[c]{@{}l@{}}Commun-\\ ication\end{tabular} & \begin{tabular}[c]{@{}l@{}}Society/\\ Government\end{tabular} & \begin{tabular}[c]{@{}l@{}}Social \\ Interaction\end{tabular} & \begin{tabular}[c]{@{}l@{}}Multi-\\ media\end{tabular} & Leisure & \begin{tabular}[c]{@{}l@{}}Health \\ Related\end{tabular} & \begin{tabular}[c]{@{}l@{}}File \\ Transfer\end{tabular} & \begin{tabular}[c]{@{}l@{}}Adult \\ Related\end{tabular} & \begin{tabular}[c]{@{}l@{}}Security \\ Threats\end{tabular} & \begin{tabular}[c]{@{}l@{}}Liability \\ Concerns\end{tabular} & \begin{tabular}[c]{@{}l@{}}Security \\ Concerns\end{tabular}  \\ \hline
Commerce                                                                & 1.00                                                   & 1.00                                                  & 1.00                                                      & 1.00                                                          & 1.00                                                          & 1.00                                                   & 1.00    & 1.00                                                      & 0.07                                              & 0.95                                            & 0.41                                               & 1.00                                                          & {$\mathbf{<10^{-5}}$}                                                                                                   \\ \hline
Technology                                                               &                                                        & 1.00                                                           & 1.00                                                       & 1.00                                                          & 1.00                                                          & 1.00                                                   & 1.00    & 1.00                                                      & 0.24                                              & 1.00                                      & 1.00                                              & 1.00                                                          & {$\mathbf{<10^{-5}}$}                                                                                                \\ \hline
\begin{tabular}[c]{@{}l@{}}Information \\ Related\end{tabular}           &                                                        &                                                                & 1.00                                             & 1.00                                                 & 1.00                                                          & 0.12                                             & 1.00    & 1.00                                                       & 1.00                                                     & 1.00                                            & 1.00                                                        & 1.00                                                          & {$\mathbf{<10^{-5}}$}                                                                                                   \\ \hline
\begin{tabular}[c]{@{}l@{}}Commun-\\ ication\end{tabular}                &                                                        &                                                                &                                                           & 1.00                                                          & 1.00                                                           & 1.00                                                    & 1.00    & 1.00                                                      & 0.22                                               & 1.00                                            & 0.74                                               & 1.00                                                          & {$\mathbf{<10^{-5}}$}                                                                                               \\ \hline
\begin{tabular}[c]{@{}l@{}}Society/\\ Government\end{tabular}            &                                                        &                                                                &                                                           &                                                               & 1.00                                                          & 1.00                                                   & 1.00    & 1.00                                                      & 0.09                                               & 1.00                                            & 0.41                                               & 1.00                                                          & {$\mathbf{<10^{-5}}$}                                                                                                  \\ \hline
\begin{tabular}[c]{@{}l@{}}Social \\ Interaction\end{tabular}            &                                                        &                                                                &                                                           &                                                               &                                                               & 1.00                                                    & 1.00    & 1.00                                                      & 1.00                                                     & 1.00                                                     & 1.00                                                        & 1.00                                                           & {$\mathbf{<10^{-5}}$}                                                                                                     \\ \hline
\begin{tabular}[c]{@{}l@{}}Multi-\\ media\end{tabular}                   &                                                        &                                                                &                                                           &                                                               &                                                               &                                                        & 1.00    & 1.00                                                       & \textbf{0.04}                                               & 1.00                                            & \textbf{0.03}                                                  & 1.00                                                          & {$\mathbf{<10^{-5}}$}                                                                                                  \\ \hline
Leisure                                                                  &                                                        &                                                                &                                                           &                                                               &                                                               &                                                        &         & 1.00                                                       & 1.00                                            & 1.00                                                     & 1.00                                               & 1.00                                                          & {$\mathbf{<10^{-5}}$}                                                                                                    \\ \hline
\begin{tabular}[c]{@{}l@{}}Health \\ Related\end{tabular}                &                                                        &                                                                &                                                           &                                                               &                                                               &                                                        &         &                                                           & 1.00                                            & 1.00                                                      & 1.00                                              & 1.00                                                          & {$\mathbf{<10^{-5}}$}                                                                                              \\ \hline
\begin{tabular}[c]{@{}l@{}}File \\ Transfer\end{tabular}                 &                                                        &                                                                &                                                           &                                                               &                                                               &                                                        &         &                                                           &                                                          & 1.00                                                     & 1.00                                                        & 1.00                                                           & {$\mathbf{<10^{-5}}$}                                                                                                        \\ \hline
\begin{tabular}[c]{@{}l@{}}Adult \\ Related\end{tabular}                 &                                                        &                                                                &                                                           &                                                               &                                                               &                                                        &         &                                                           &                                                          &                                                          & 1.00                                               & 1.00                                                          & {$\mathbf{<10^{-5}}$}                                                                                                     \\ \hline
\begin{tabular}[c]{@{}l@{}}Security \\ Threats\end{tabular}              &                                                        &                                                                &                                                           &                                                               &                                                               &                                                        &         &                                                           &                                                          &                                                          &                                                             & 1.00                                                          & {$\mathbf{<10^{-5}}$}                                                                                                     \\ \hline
\begin{tabular}[c]{@{}l@{}}Liability \\ Concerns\end{tabular}            &                                                        &                                                                &                                                           &                                                               &                                                               &                                                        &         &                                                           &                                                          &                                                          &                                                             &                                                               & {$\mathbf{<10^{-5}}$}                                                                                                     \\ \hline

\end{tabular}
\end{adjustbox}

%% file: figure_delta_hours_pday_puser.tex
\subfloat[][User 6128]{\includegraphics[width=.2\textwidth]{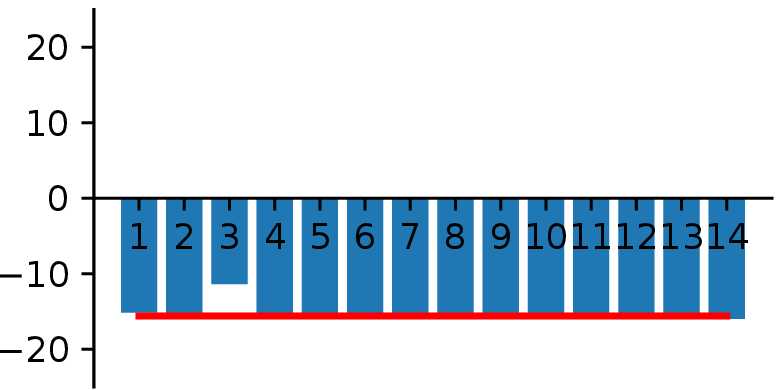}}
	\subfloat[][User 0057]{\includegraphics[width=.2\textwidth]{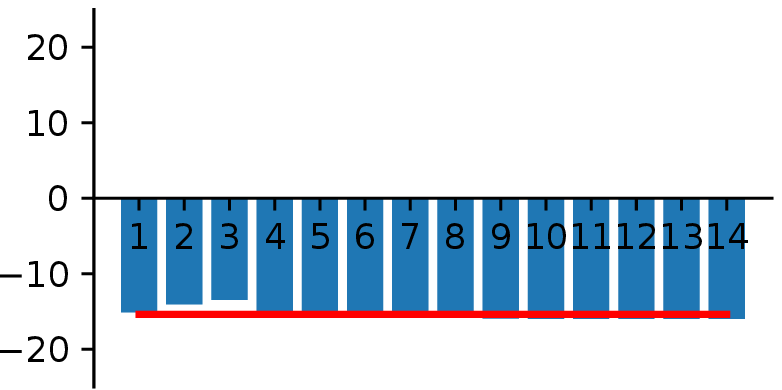}}
	\subfloat[][User 1727]{\includegraphics[width=.2\textwidth]{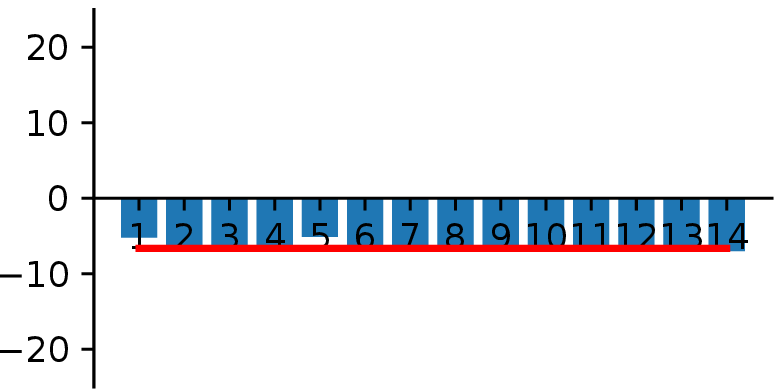}}
	\subfloat[][User 6189]{\includegraphics[width=.2\textwidth]{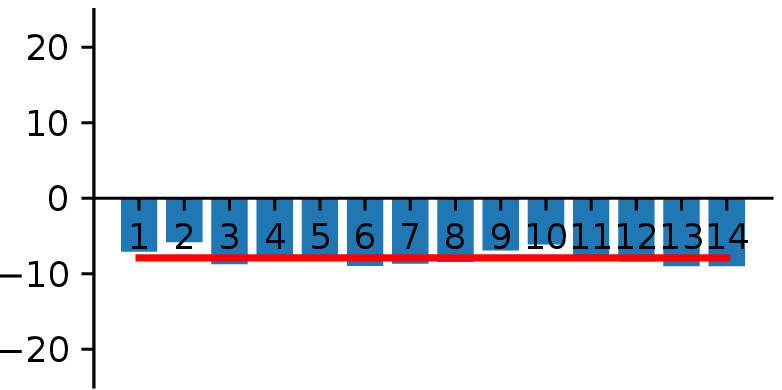}}
	\subfloat[][User 4601]{\includegraphics[width=.2\textwidth]{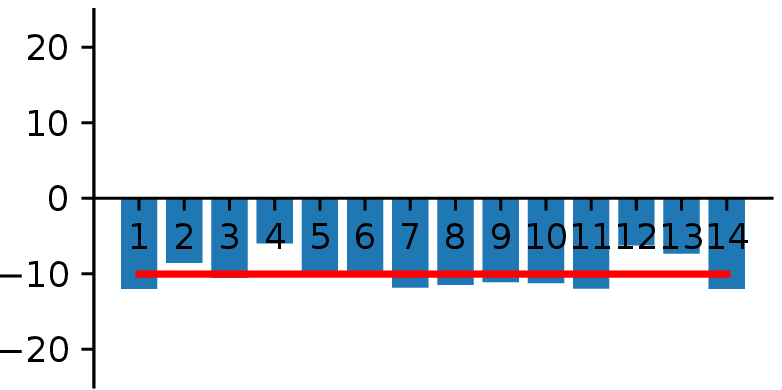}}
	\\
	\subfloat[][User 0366]{\includegraphics[width=.2\textwidth]{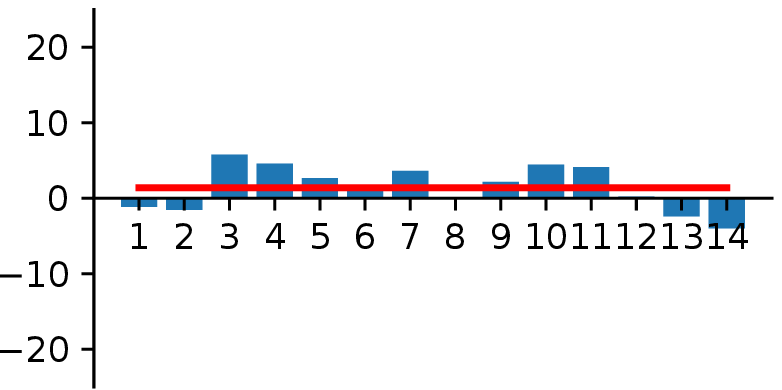}}
	\subfloat[][User 0824]{\includegraphics[width=.2\textwidth]{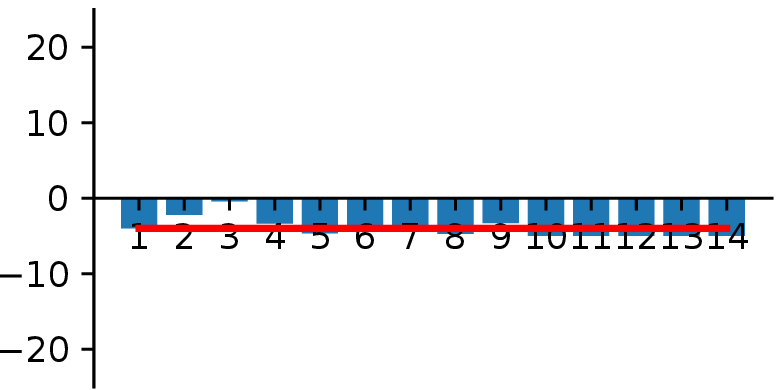}}
	\subfloat[][User 0455]{\includegraphics[width=.2\textwidth]{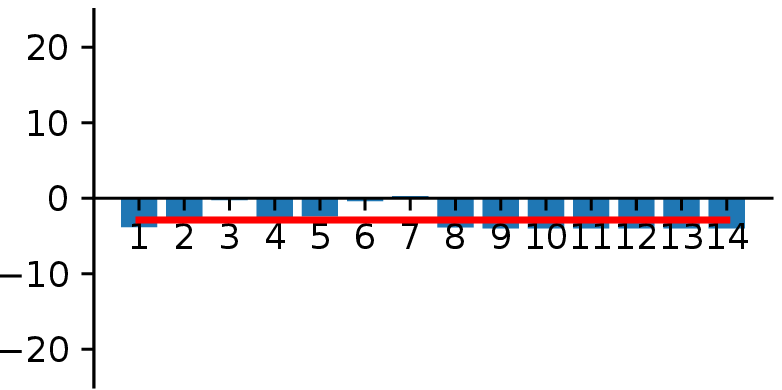}}
	\subfloat[][User 1287]{\includegraphics[width=.2\textwidth]{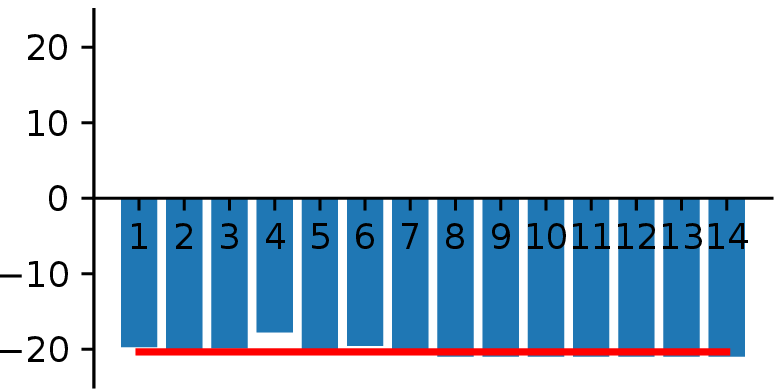}}
	\subfloat[][User 2092]{\includegraphics[width=.2\textwidth]{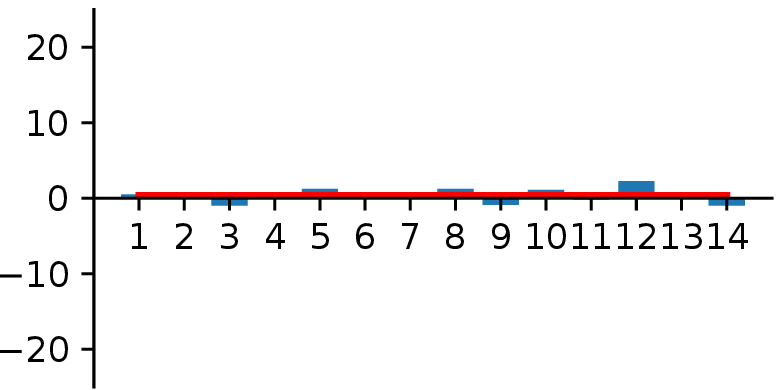}}
	\\
	\subfloat[][User 1803]{\includegraphics[width=.2\textwidth]{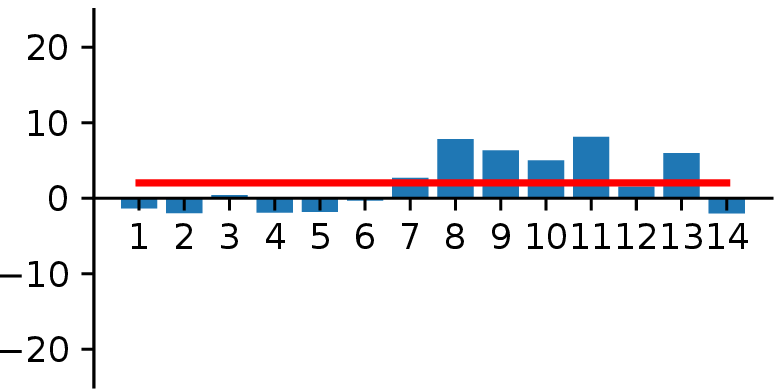}}
	\subfloat[][User 7038]{\includegraphics[width=.2\textwidth]{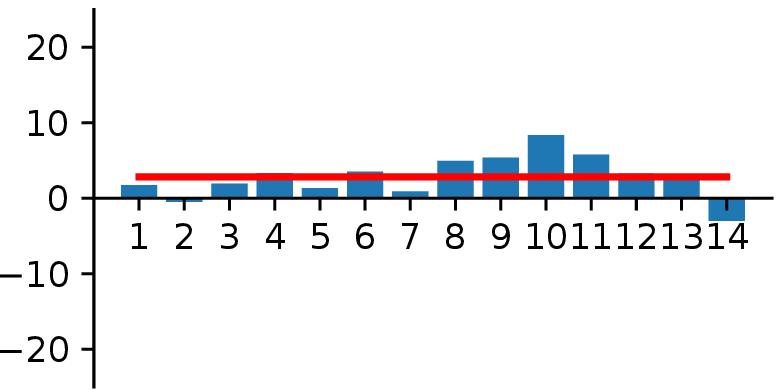}}
	\subfloat[][User 6566]{\includegraphics[width=.2\textwidth]{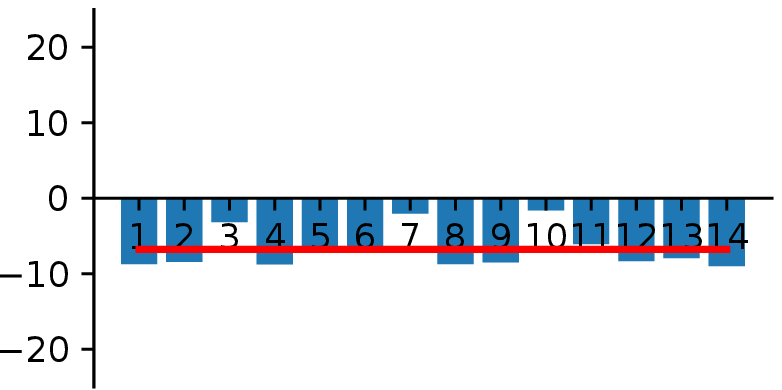}}
	\subfloat[][User 4940]{\includegraphics[width=.2\textwidth]{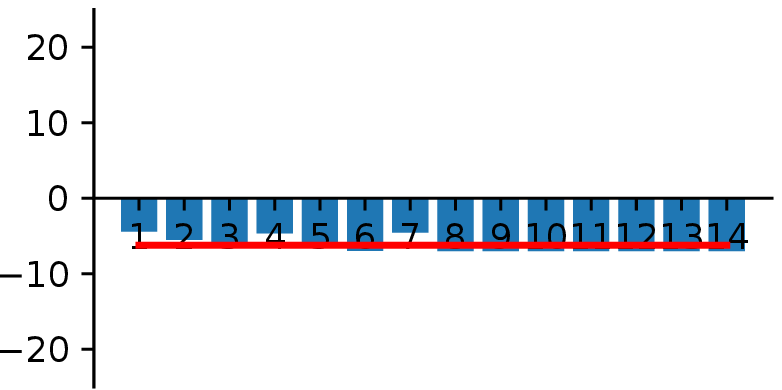}}
	\subfloat[][User 1415]{\includegraphics[width=.2\textwidth]{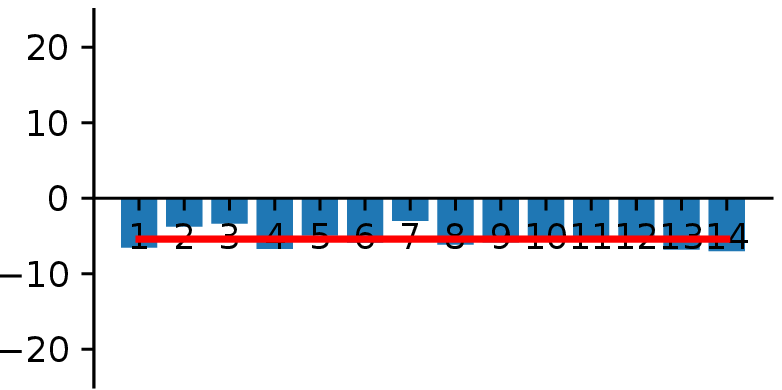}}
	\\
	\subfloat[][User 0330]{\includegraphics[width=.2\textwidth]{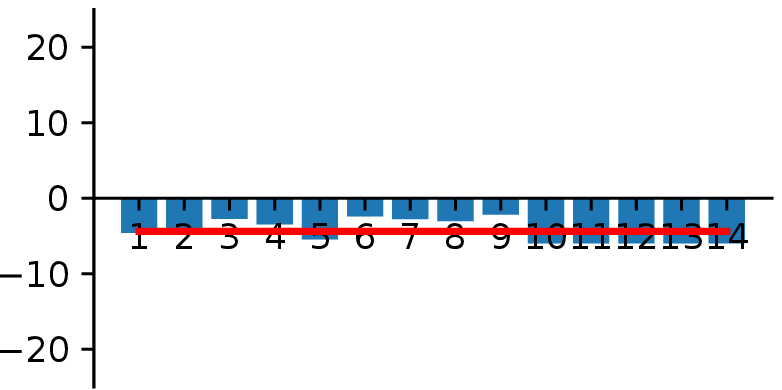}}
	\subfloat[][User 1842]{\includegraphics[width=.2\textwidth]{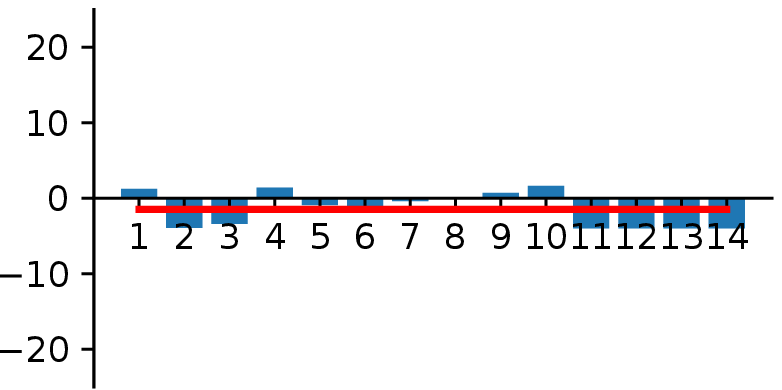}}
	\subfloat[][User 2254]{\includegraphics[width=.2\textwidth]{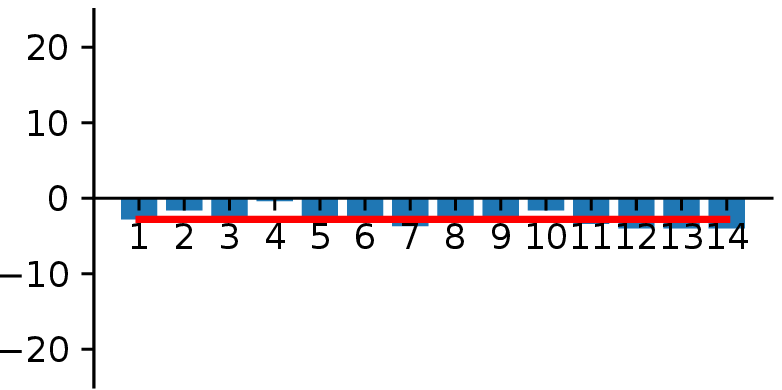}}
	\subfloat[][User 6966]{\includegraphics[width=.2\textwidth]{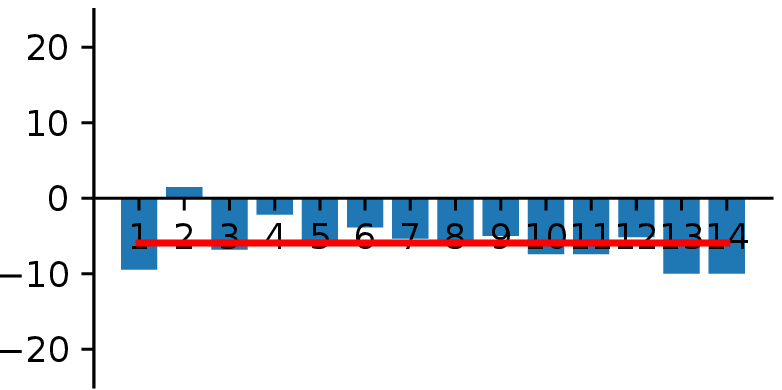}}
	\subfloat[][User 0689]{\includegraphics[width=.2\textwidth]{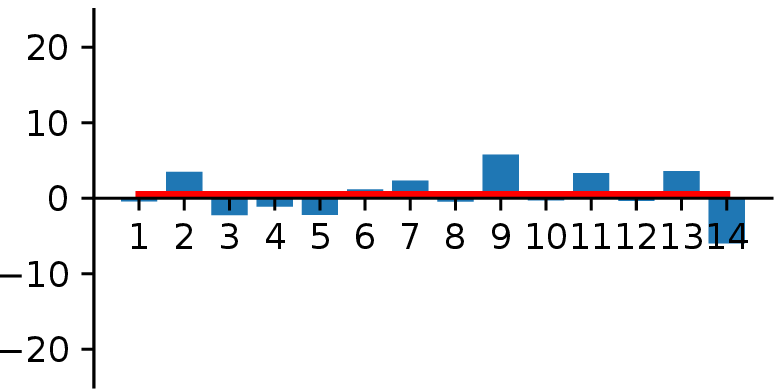}}
	\\
	\subfloat[][User 0689]{\includegraphics[width=.2\textwidth]{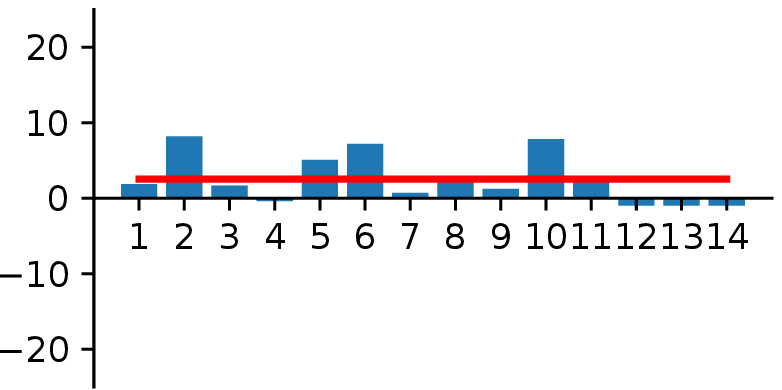}}
	\subfloat[][User 4174]{\includegraphics[width=.2\textwidth]{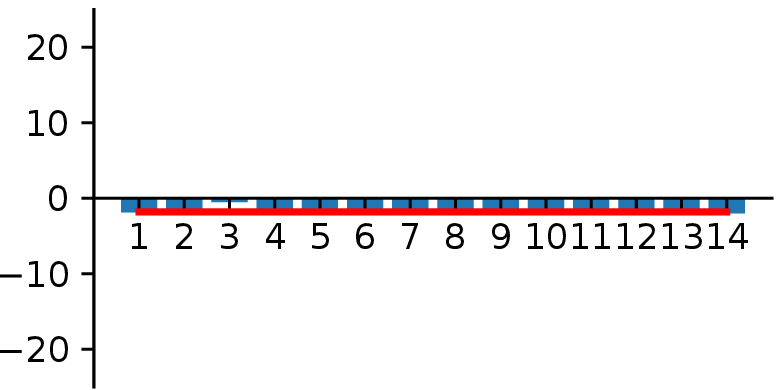}}
	\subfloat[][User 4077]{\includegraphics[width=.2\textwidth]{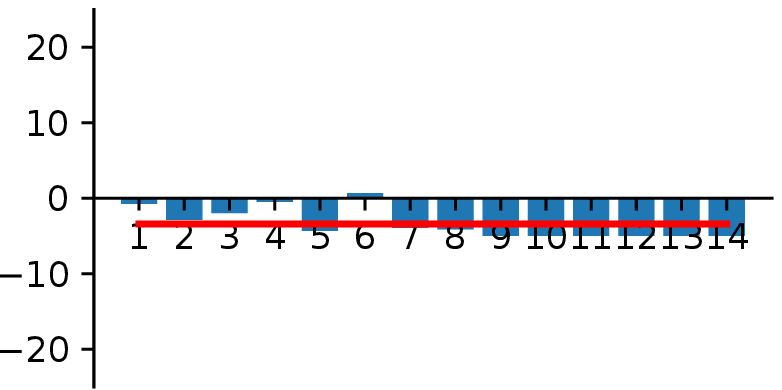}}
	\subfloat[][User 3254]{\includegraphics[width=.2\textwidth]{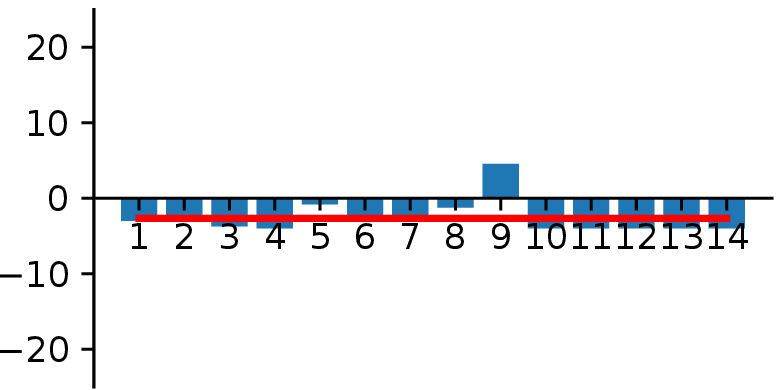}}
	\subfloat[][User 6919]{\includegraphics[width=.2\textwidth]{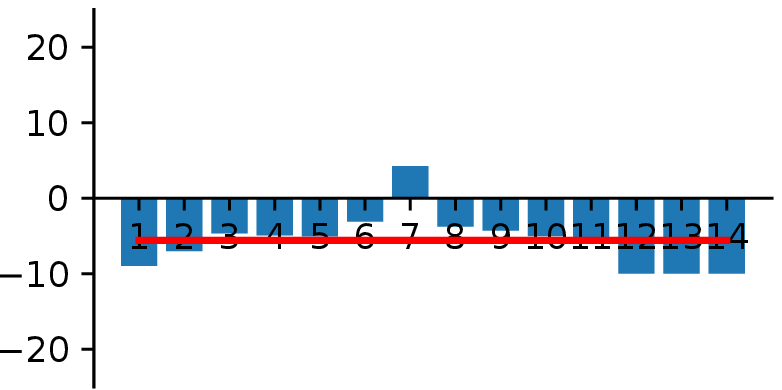}}
	\\
	\subfloat[][User 7654]{\includegraphics[width=.2\textwidth]{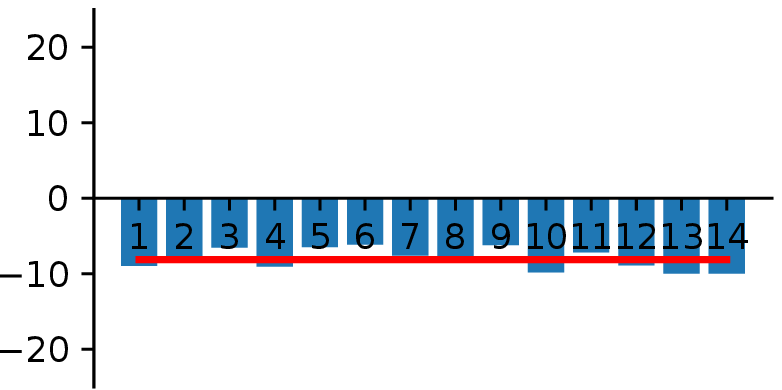}}
	\subfloat[][User 3254]{\includegraphics[width=.2\textwidth]{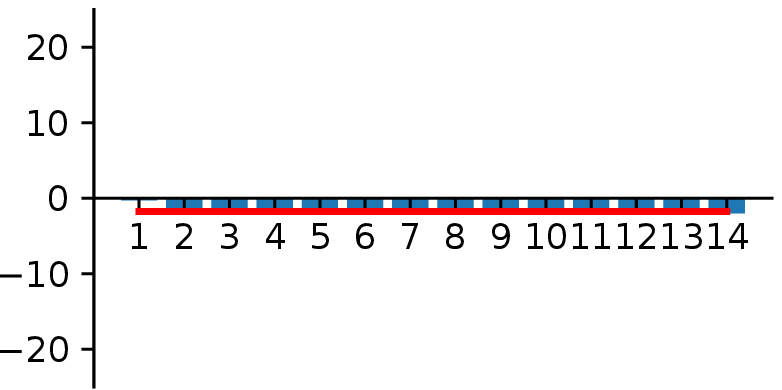}}
	\subfloat[][User 4077]{\includegraphics[width=.2\textwidth]{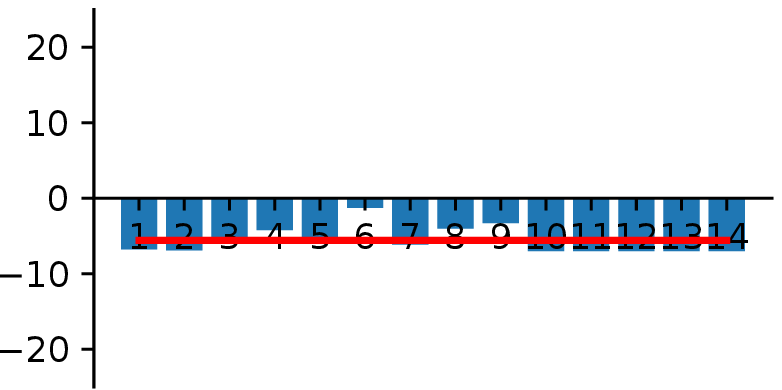}}
	\subfloat[][User 5400]{\includegraphics[width=.2\textwidth]{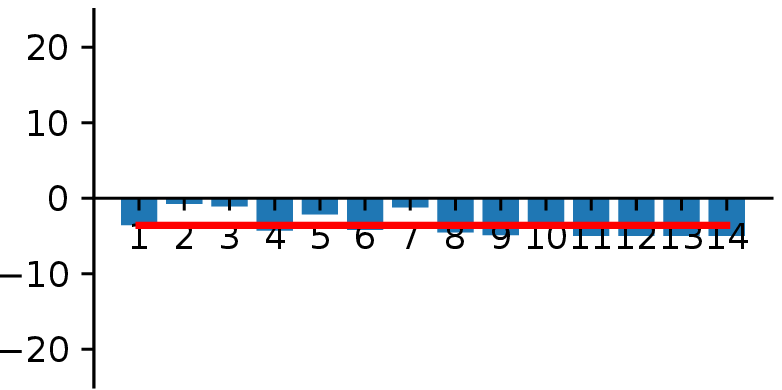}}
	\subfloat[][User 4174]{\includegraphics[width=.2\textwidth]{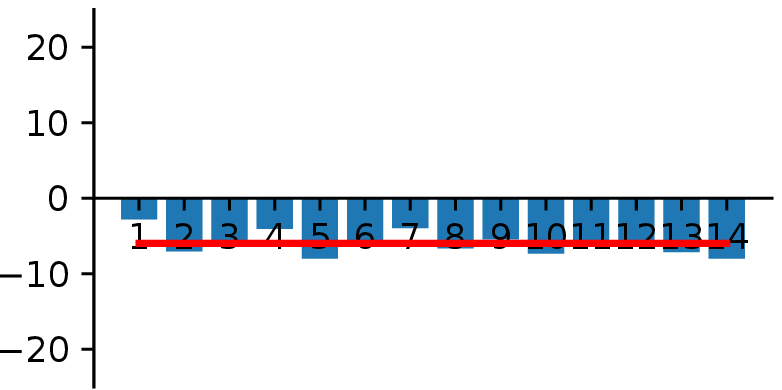}}
	\\
	\subfloat[][User 6744]{\includegraphics[width=.2\textwidth]{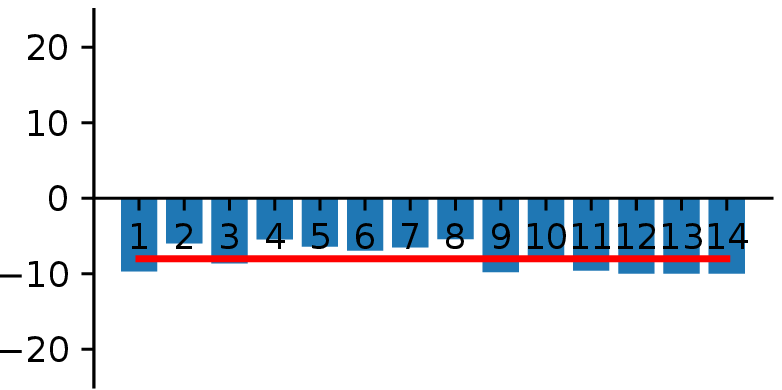}}
	\\

%% file: main.bbl
\begin{thebibliography}{26}
\providecommand{\natexlab}[1]{#1}
\providecommand{\url}[1]{\texttt{#1}}
\expandafter\ifx\csname urlstyle\endcsname\relax
  \providecommand{\doi}[1]{doi: #1}\else
  \providecommand{\doi}{doi: \begingroup \urlstyle{rm}\Url}\fi

\bibitem[Abramson and Gore(2013)]{abramson2013associative}
Myriam Abramson and Shantanu Gore.
\newblock Associative patterns of web browsing behavior.
\newblock In \emph{AAAI Fall Symposia}, 2013.

\bibitem[Amazon()]{alexa}
Amazon.
\newblock Alexa top sites by category.
\newblock \url{ https://www.alexa.com/topsites/categories}.
\newblock Accessed: 6/29/2018.

\bibitem[Andreoni and Bernheim(2009)]{AB09}
James Andreoni and B.~Douglas Bernheim.
\newblock Social image and the 50-50 norm: A theoretical and experimental
  analysis of audience effects.
\newblock \emph{Econometrica}, 77\penalty0 (5):\penalty0 1607--1636, 2009.

\bibitem[Ariely et~al.(2009)Ariely, Bracha, and Meier]{ABM09}
Dan Ariely, Anat Bracha, and Stephan Meier.
\newblock Doing good or doing well? {I}mage motivation and monetary incentives
  in behaving prosocially.
\newblock \emph{American Economic Review}, 99\penalty0 (1):\penalty0 544--555,
  2009.

\bibitem[BroadbandSearch(2021)]{broad}
BroadbandSearch.
\newblock Mobile vs. desktop internet usage \hspace{9cm}.
\newblock
  \url{https://www.broadbandsearch.net/blog/mobile-desktop-internet-usage-statistics},
  2021.
\newblock Accessed: 5/4/2021.

\bibitem[Calafiore and Damianov(2011)]{calafiore2011effect}
Pablo Calafiore and Damian~S Damianov.
\newblock The effect of time spent online on student achievement in online
  economics and finance courses.
\newblock \emph{The Journal of Economic Education}, 42\penalty0 (3):\penalty0
  209--223, 2011.

\bibitem[Cao and Caverlee(2014)]{cao2014behavioral}
Cheng Cao and James Caverlee.
\newblock Behavioral detection of spam {URL} sharing: {P}osting patterns versus
  click patterns.
\newblock In \emph{Proceedings of the 2014 IEEE/ACM International Conference on
  Advances in Social Networks Analysis and Mining}, ASONAM '14, pages 138--141,
  2014.

\bibitem[Cao and Caverlee(2015)]{cao2015detecting}
Cheng Cao and James Caverlee.
\newblock Detecting spam {URLs} in social media via behavioral analysis.
\newblock In \emph{Advances in Information Retrieval}, pages 703--714. Springer
  International Publishing, 2015.

\bibitem[Dong et~al.(2008)Dong, Clark, and Jacob]{dong2008user}
Xun Dong, John~A. Clark, and Jeremy~L. Jacob.
\newblock User behaviour based phishing websites detection.
\newblock In \emph{Proceedings of the 2008 International Multiconference on
  Computer Science and Information Technology}, pages 783--790, 2008.

\bibitem[Duffy and Chan(2019)]{duffy2019you}
Brooke~Erin Duffy and Ngai~Keung Chan.
\newblock {“You never really know who’s looking”: I}magined surveillance
  across social media platforms.
\newblock \emph{New Media \& Society}, 21\penalty0 (1):\penalty0 119--138,
  2019.

\bibitem[Efron and Tibshirani(1986)]{efron1986bootstrap}
Bradley Efron and Robert Tibshirani.
\newblock Bootstrap methods for standard errors, confidence intervals, and
  other measures of statistical accuracy.
\newblock \emph{Statistical science}, pages 54--75, 1986.

\bibitem[Ellison et~al.(2007)Ellison, Steinfield, and
  Lampe]{ellison2007benefits}
Nicole~B. Ellison, Charles Steinfield, and Cliff Lampe.
\newblock {The benefits of Facebook “friends:” S}ocial capital and college
  students’ use of online social network sites.
\newblock \emph{Journal of Computer-Mediated Communication}, 12\penalty0
  (4):\penalty0 1143--1168, 2007.

\bibitem[Ernala et~al.(2020)Ernala, Burke, Leavitt, and
  Ellison]{10.1145/3313831.3376435}
Sindhu~Kiranmai Ernala, Moira Burke, Alex Leavitt, and Nicole~B. Ellison.
\newblock How well do people report time spent on {F}acebook? {A}n evaluation
  of established survey questions with recommendations.
\newblock In \emph{Proceedings of the 2020 CHI Conference on Human Factors in
  Computing Systems}, CHI '20, pages 1--14, 2020.

\bibitem[Goel et~al.(2012)Goel, Hofman, and Sirer]{goel2012does}
Sharad Goel, Jake Hofman, and M~Sirer.
\newblock Who does what on the web: A large-scale study of browsing behavior.
\newblock In \emph{Proceedings of the International AAAI Conference on Web and
  Social Media}, ICWSM '12, pages 130--137, 2012.

\bibitem[Groves et~al.(2011)Groves, Fowler~Jr, Couper, Lepkowski, Singer, and
  Tourangeau]{groves2011survey}
Robert~M Groves, Floyd~J Fowler~Jr, Mick~P Couper, James~M Lepkowski, Eleanor
  Singer, and Roger Tourangeau.
\newblock \emph{Survey Methodology}, volume 561.
\newblock John Wiley \& Sons, 2011.

\bibitem[Hargittai(2010)]{hargittai2010digital}
Eszter Hargittai.
\newblock Digital natives? {V}ariation in internet skills and uses among
  members of the ``net generation''.
\newblock \emph{Sociological Inquiry}, 80\penalty0 (1):\penalty0 92--113, 2010.

\bibitem[Hoffman et~al.(1996)Hoffman, McCabe, and Smith]{HMS96}
Elizabeth Hoffman, Kevin McCabe, and Vernon Smith.
\newblock Social distance and other-regarding behavior in dictator games.
\newblock \emph{American Economic Review}, 86\penalty0 (3):\penalty0 653--660,
  1996.

\bibitem[Hu et~al.(2007)Hu, Zeng, Li, Niu, and Chen]{hu2007demographic}
Jian Hu, Hua-Jun Zeng, Hua Li, Cheng Niu, and Zheng Chen.
\newblock Demographic prediction based on user's browsing behavior.
\newblock In \emph{Proceedings of the 16th International Conference on World
  Wide Web}, WWW '07, pages 151--160, 2007.

\bibitem[Kumar and Tomkins(2010)]{kumar2010characterization}
Ravi Kumar and Andrew Tomkins.
\newblock A characterization of online browsing behavior.
\newblock In \emph{Proceedings of the 19th International Conference on World
  Wide Web}, WWW '10, pages 561--570, 2010.

\bibitem[Ostrom et~al.(1992)Ostrom, Walker, and Gardner]{OWG92}
Elinor Ostrom, James Walker, and Roy Gardner.
\newblock Covenants with and without a sword: Self-governance is possible.
\newblock \emph{The American Political Science Review}, 86\penalty0
  (2):\penalty0 404--417, 1992.

\bibitem[Redmiles et~al.(2018)Redmiles, Zhu, Kross, Kuchhal, Dumitras, and
  Mazurek]{redmiles2018asking}
Elissa~M Redmiles, Ziyun Zhu, Sean Kross, Dhruv Kuchhal, Tudor Dumitras, and
  Michelle~L Mazurek.
\newblock Asking for a friend: Evaluating response biases in security user
  studies.
\newblock In \emph{Proceedings of the 2018 ACM SIGSAC Conference on Computer
  and Communications Security}, CCS '18, pages 1238--1255, 2018.

\bibitem[Rege and Telle(2004)]{RT04}
Mari Rege and Kjetil Telle.
\newblock The impact of social approval and framing on cooperation in public
  good situations.
\newblock \emph{Journal of Public Economics}, 88\penalty0 (7):\penalty0
  1625--1644, 2004.

\bibitem[Symantec()]{Symantec}
Symantec.
\newblock Webpulse site review request.
\newblock \url{https://sitereview.bluecoat.com/}.
\newblock Accessed: 9/30/2020.

\bibitem[Thomas et~al.(2011)Thomas, Grier, Ma, Paxson, and
  Song]{thomas2011design}
Kurt Thomas, Chris Grier, Justin Ma, Vern Paxson, and Dawn Song.
\newblock Design and evaluation of a real-time {URL} spam filtering service.
\newblock In \emph{Proceedings of the 2011 IEEE Symposium on Security and
  Privacy}, S\&P '11, pages 447--462, 2011.

\bibitem[Wash et~al.(2017)Wash, Rader, and Fennell]{wash2017can}
Rick Wash, Emilee Rader, and Chris Fennell.
\newblock Can people self-report security accurately? {A}greement between
  self-report and behavioral measures.
\newblock In \emph{Proceedings of the 2017 CHI Conference on Human Factors in
  Computing Systems}, CHI '17, pages 2228--2232, 2017.

\bibitem[Williams et~al.(2017)Williams, Jenkins, Valacich, and
  Byrd]{williams2017measuring}
Parker~A Williams, Jeffrey Jenkins, Joseph Valacich, and Michael~D Byrd.
\newblock Measuring actual behaviors in {HCI} research--{A} call to action and
  an example.
\newblock \emph{Transactions on Human-Computer Interaction}, 9\penalty0
  (4):\penalty0 339--352, 2017.

\end{thebibliography}
